\documentclass[a4paper,11pt]{article}
\usepackage{jcappub} % for details on the use of the package, please see the JINST-author-manual
\usepackage{booktabs}

\usepackage{mathtools}
\usepackage{lineno}
\arxivnumber{1234.56789} % Only if you have one
\title{\boldmath Into the Gompverse: A robust Gompertzian reionization model for CMB analyses}

\newcommand{\As}{A_\mathrm{s}}
\newcommand{\ns}{n_\mathrm{s}}
\newcommand{\Omegam}{\Omega_\mathrm{m}}
\newcommand{\Omegac}{\Omega_\mathrm{c}}
\newcommand{\Omegab}{\Omega_\mathrm{b}}

\newcommand{\zetaUV}{\zeta_\mathrm{UV}}
\newcommand{\Tvir}{T_\mathrm{vir}}
\newcommand{\LX}{L_\mathrm{X}}
\newcommand{\HI}{\mathrm{HI}}

\newcommand{\re}{\mathrm{re}}
\newcommand{\reio}{\mathrm{reio}}

\author[a]{Paulo Montero-Camacho,}
\author[a]{Yin Li,}
\author[b]{Marco Barquero-Hern\'andez,}
\author[c, d]{Pablo Renard,}
\author[a]{Yang Wang,}
\author[e]{and Xiao-Dong Li}
\affiliation[a]{Department of Strategic and Advanced Interdisciplinary Research, Pengcheng Laboratory,\\
Shenzhen, Guangdong 518066, China}
\affiliation[b]{Escuela de Física, Universidad de Costa Rica, 11501 San José, Costa Rica}
\affiliation[c]{Institute of Space Sciences (ICE, CSIC), Campus UAB, Carrer de Can Magrans, s/n, 08193 Barcelona, Spain}
\affiliation[d]{Institut d'Estudis Espacials de Catalunya (IEEC), E-08860 Castelldefels (Barcelona), Spain}
\affiliation[e]{School of Physics and Astronomy, Sun Yat-Sen University, Zhuhai 519082, People's Republic of China}

\emailAdd{pmontero@pcl.ac.cn}
\emailAdd{eelregit@gmail.com}

\abstract{Cosmic reionization is driven by the formation of sources of ultraviolet photons, and hence it is an intrinsically asymmetric process, where its earlier stages occur at a slower pace relative to its later stages. Yet most modern cosmic microwave background (CMB) analyses rely on a hyperbolic tangent template, i.e. a symmetric sigmoid, that is not well suited for joint fitting of CMB and reionization observations. In this work, we introduce a physically motivated Gompertzian reionization model with three astrophysical (\emph{nuisance}) parameters, designed to enable joint analyses of CMB and reionization data and to be applicable to a wide range of datasets and cosmological models. This robust Gompertzian model leverages the connection between cosmology and reionization, typically ignored in standard CMB analyses, to demote the optical depth ($\tau_\reio$) to derived parameter, reducing its uncertainty by approximately a factor of three compared to the conventional $\tanh$ prescription. The $\tau_\reio$ improvement enables tighter constraints on the sum of the neutrino masses, revealing potential tension with neutrino oscillation experiments even after accounting for the known relaxation of neutrino mass bounds in $w_0w_a$CDM models -- a tension that is partially obscured by the conventional treatment of reionization. In addition, the inferred constraints on the astrophysical parameters governing reionization naturally synergize with current and upcoming 21 cm experiments, providing physically informed parameter ranges for future 21 cm studies. }


\begin{document}
\maketitle
\flushbottom

\section{Introduction}
\label{sec:intro}

Cosmic reionization of hydrogen, and its accompanying first ionization of helium, is a complex phase transition that the universe goes through around redshift $\mathcal{O}(10)$ \cite{2020A&A...641A...6P}. In early stages of reionization, due to the the scarcity of ultraviolet photon sources capable of ionizing hydrogen, the intergalactic medium (IGM) is mostly neutral. As more sources of ionizing photons form via gravitational instability, the IGM starts to ionize, leading to an asymmetric phase transition where the first half of reionization takes a longer time to complete than the latter half. While reionization is primarily an astrophysical process, its importance extends into cosmology. This is because modeling the free electrons from reionization is essential for CMB analysis, given the Thomson scattering between these electrons and CMB photons.

Most recent CMB analyses rely on modeling the fraction of free electrons during and after reionization with a hyperbolic tangent. This symmetric model was introduced in \cite{2008PhRvD..78b3002L} and has become standard in the data analysis of state-of-the-art CMB experiments such as Planck (e.g., \cite{2020A&A...641A...6P}) and the Atacama Cosmology Telescope (e.g., \cite{2025arXiv250314454C}), achieving powerful constraints on cosmological parameters. Among the six free parameters in the $\Lambda$CDM model, only the optical depth to reionization ($\tau_\mathrm{reio}$) remains poorly constrained $\gtrsim 1\%$. See Figure 1 of \cite{2024arXiv240513680M} for a compilation of recent representative constraints on $\tau_\mathrm{reio}$.

An accurate measurement of the optical depth would not only yield information about reionization \cite{2025arXiv250515899C,2025arXiv250413254I} but also alleviate degeneracies with other cosmological parameters, such as degeneracies with both the amplitude of the primordial scalar power spectrum ($\As$) and the tensor-to-scalar ratio \cite{2020A&A...644A..32N}, thus leading to tighter constraints on cosmological models. Furthermore, recent developments \cite{2025arXiv250416932S,2025arXiv250421813J} indicate that the apparent preference for negative sum of the neutrino masses and potentially the preference for evolving dark energy could be alleviated by a higher optical depth ($\tau_\mathrm{reio} \sim 0.09$) in significant tension with Planck low-$\ell$ measurements ($\tau_\mathrm{reio} \sim 0.06$). We caution that such large values of $\tau_\mathrm{reio}$ are likely to lead to conflicts with observational constraints on the timeline of reionization.

While the $\tanh$ model has been highly effective for modern CMB analyses, the best fits from CMB data produce inadequate fits to astrophysical constraints on the timeline of reionization \cite{2024arXiv240513680M,2025PhRvD.111d3532P}. In particular, \cite{2025PhRvD.111d3532P} reported decisive evidence -- using Bayesian evidence ratio -- for an asymmetric reionization history from a joint analysis of CMB, ultraviolet luminosity function, and constraints on the timeline of reionization. Furthermore, future CMB experiments are anticipated to reach the cosmic variance limit\footnote{This limit assumes the standard hyperbolic tangent prescription and therefore ignores the coupling between cosmological parameters and reionization.} for $\tau_\reio$ \cite{2023PTEP.2023d2F01L}, which will likely exacerbate this discrepancy with astrophysical constraints on the timeline of reionization. Likewise, future CMB missions will be able to probe the patchiness of reionization \cite{2018JCAP...05..014R} and achieve better indirect constraints on properties of the reionization timeline, such as its asymmetry and duration \cite{2018JCAP...09..016H, 2020ApJ...889..130W}.

In this work, we present a robust reionization model capable of jointly fitting CMB and reionization constraints. Our model is based on the Gompertzian reionization of \cite{2024arXiv240513680M}, which yields tighter cosmological constraints than conventional analyses due to its ability to fit reionization observations. We supplement this ability with a mapping between cosmology, astrophysics, and reionization timeline based on {\sc 21cmFAST}\footnote{This semi-numerical code is at the forefront of 21 cm analyses, e.g. \cite{2025arXiv251121289T}.} simulations \cite{2011MNRAS.411..955M} and constructed with symbolic regression \cite{2013MNRAS.431.2371G}. Consequently, our model produces tighter cosmological constraints than conventional analyses while being fully interpretable, as the model is composed of analytical expressions where cosmological parameters are explicitly included.

The rest of this paper is organized as follows. We introduce our new reionization model for CMB analyses in \S\ref{sec:gomp}. In \S\ref{sec:meth}, we describe the different data sets, the different Boltzmann codes, and the sampler used in our Bayesian inference of cosmological parameters. We validate and compare the robust Gompertzian model with other reionization models in \S\ref{sec:val}. In \S\ref{sec:results}, we present our main results, the cosmological information gained while utilizing  our new model for different data sets and cosmologies. Finally, we summarize our findings in \S\ref{sec:sum}. In addition, Appendix \ref{app:trends} displays the different dependence of our model with cosmological and astrophysical parameters, while Appendix \ref{app:tanh} reinforces the inability of the conventional hyperbolic tangent reionization model to simultaneously fit CMB and reionization information.

\section{A Robust Gompertzian reionization model}
\label{sec:gomp}
While cosmic reionization is a complex phase transition, accumulated knowledge now clearly indicates that it must be an asymmetric process with a slow start and rapid evolution in the later stages. Yet, state-of-the-art analyses of CMB data commonly utilize symmetric sigmoids, which are known to be inadequate for reproducing reionization data. The main purpose of this paper is to present a new reionization model for CMB analyses that not only aligns with our theoretical expectations and observational constraints of reionization, but is also more robust than the simple model introduced in \cite{2024arXiv240513680M}. In their model, the uncertainty corresponding to the reionization modeling is essentially stored in a single free parameter, the ionization efficiency ($\zeta_\mathrm{UV}$), which quantifies the difficulty for ionizing photons to escape their parent galaxies and contribute to the ionization of the IGM. Note that larger values of $\zetaUV$ result in earlier reionization scenarios.

Naturally, another parameter of interest for reionization modeling is the difficulty for those parent galaxies to form, which here we will parametrize with their virial temperature $T_\mathrm{vir}$, a commonly used strategy (see e.g., \cite{2019PhRvD.100f3538M,2023MNRAS.525.6097S, 2024A&A...686A.212H}). In contrast to $\zetaUV$, higher values of $\Tvir$ lead to delays in the timeline of reionization because it is more difficult for the source galaxies to form \cite{2020MNRAS.499.1640M}. Similarly, this threshold also influences the speed of reionization since an increase in the number of sources significantly impacts the rate of ionizing photons.

In addition to ultraviolet radiation, X-ray heating from sources like high-redshift binaries \cite{2014ApJ...791..110X} could contribute to early IGM ionization, becoming another important feature of cosmic reionization \cite{2013MNRAS.431..621M,2024MNRAS.529.3666M}.
The degree of heating, its sources, and the impact on the neutral hydrogen fraction $x_{\rm HI}$ are uncertain. X-ray heating is often parametrized using two free parameters \cite{2018IAUS..333...18G}: $E_0$, the energy threshold for the lowest energy X-ray photons not absorbed by galaxies, and the integrated soft-band luminosity ($L_{\mathrm{X}<2\, \mathrm{keV}}/\mathrm{SFR} \equiv L_\mathrm{X}$). The former parameter quantifies the impact that interstellar medium absorption has on the X-ray spectral energy density, while the latter parameter is a proxy for the expected emission of X-ray photons produced during the cosmic dawn. While X-ray heating can substantially impact the 21 cm signal \cite{2013MNRAS.431..621M}, reasonable values of $E_0$ would have negligible impact on the timeline of reionization \cite{2020MNRAS.499.1640M}. In contrast, inefficient X-ray heating, i.e. smaller values of $\LX$ will slightly delay reionization.

Given their role in the evolution of the neutral hydrogen fraction, our robust Gompertzian reionization model will have three astrophysical free parameters $\zetaUV$, $\Tvir$, and $\LX$ that must be marginalized over for cosmological constraints using CMB data.

We base our model on the Gompertzian shape presented in \cite{2024arXiv240513680M}. This asymmetric curve easily matches observational constraints on the time of reionization and showcases an early low start with rapid acceleration past the midpoint. The neutral hydrogen evolution in our Robust Gompertzian model is given by 
\begin{align}
    &x_\HI(\tilde{a}) = \exp{\{-\exp[P_5(\tilde{a})]\}} \, , \label{eq:gomp} \\
    &P_5(\tilde{a}) = \sum_{m = 0}^{5} c_m \ln^m \tilde{a} \, , \\
    &\boldsymbol{c} = \{0, 1, 0.1252, 0.03533, 0.002203, 0.000007483 \}\, , \\
    &\tilde{a}(a;\boldsymbol{\theta}) = \left[\frac{a}{\alpha(\boldsymbol{\theta})} \right]^{\beta(\boldsymbol{\theta})}\, \label{eq:a_rescale} ,
\end{align}
where $a$ is the scale factor, $\boldsymbol{\theta}$ denotes 8 astrophysical and cosmological parameters, $\alpha(\boldsymbol{\theta})$ is the power-law pivot, and $\beta(\boldsymbol{\theta})$ a rescaling tilt. Furthermore, the 5th-degree polynomial is introduced to account for residuals from the Gompertz shape. Following \cite{2024arXiv240513680M}, both the $c_m$ and the parameter dependencies of the pivot and tilt stem from a suite of 1024 Sobol samples of {\sc 21cmFAST}\cite{2011MNRAS.411..955M,2020JOSS....5.2582M} simulations.

Our {\sc 21cmFAST} simulations have a 300 comoving Mpc box size with $768^3$ and $256^3$ cells for the matter and HI fields, respectively. Besides cosmological parameters and the three astrophysical parameters, most options are kept to their default values. The simulations encompass the following parameter ranges:
\begin{alignat}{4}
\label{eq:prior}
\sigma_8 &\in (0.74, 0.90), &\quad
\ns &\in (0.92, 1.00), &\quad
h &\in (0.61, 0.73), &\quad \nonumber
\Omegab &\in (0.04, 0.06), \\
\Omegam &\in (0.24, 0.40), &\quad
\zetaUV &\in (20, 35), &\quad 
\log(\Tvir) &\in (4.3, 4.7), &\quad
\log(\LX) &\in (39, 41).
\end{alignat}

As in \cite{2024arXiv240513680M}, we split the samples into two sets. We have 512 simulations that sample uniformly the parameter range defined in Eq.~(\ref{eq:prior}). Furthermore, we have extra 512 simulations that sample uniformly in an inner core region closer to the 5$\sigma$ range from the Planck PR3 cosmological constraints \cite{2020A&A...641A...6P}. These 512 inner simulations refine the cosmological dependence in the actual range of interest and are necessary since most of the 8D hypervolume lies near its surface.

With Eqs.~(\ref{eq:gomp} -- \ref{eq:a_rescale}) and our 1024 $x_{\rm HI}(z)$ samples we can establish the mapping between cosmological parameters and reionization through the pivot and rescaling of each simulation. In total, we have $4 + 2 \times 512$ parameters to determine the joint fit. The first four parameters correspond to the coefficients $c_m$\footnote{We fix the $c_0 = 0$ and $c_1 = 1$ since they are fully degenerate with $\ln \alpha$ and $\beta$.} while the fitted $\alpha$ and $\beta$ are then used as our targets for symbolic regression.

We use symbolic regression \cite{2020arXiv200611287C} -- via the {\sc PySR} package \cite{2023arXiv230501582C} -- to search a vast function space for the expressions of $\ln \alpha$ and $\beta$ as functions of cosmology and astrophysical parameters. Our search space encompasses the set of expressions composed of the operators $+$, $-$, $\cdot$, $/$, power, $\exp$, and $\ln$. We point readers interested in the specific details of the implementation, including the use of the Pareto front for model selection, to the methodology described in \cite{2024arXiv240513680M} and to our repository\footnote{\url{https://github.com/eelregit/5par/tree/master/robust_gomp}} for intermediate results. We obtain the following expressions
\begin{align}
    &\ln \alpha(\boldsymbol{\theta}) = 0.385\left(\Tvir - \sigma_8 \ln (\Tvir^{-40.065 + \LX} + \zetaUV + e^{\sigma_8}) \right) - e^{\{\ns - \Omegab^{\Omegam} \}} - 0.863^{-\ns} h^{\Omegam^{0.792}} \label{eq:pivot} \, ,\\
    &\beta(\boldsymbol{\theta}) = \left( \frac{\zetaUV}{h} \right)^{0.499} - \sigma_8^{-1}\left( \LX (\Omegab - 0.209) + 6.081 - \Omegam (\Tvir - e^{\ns^{2.803}})\right) \label{eq:tilt}\, .
\end{align}

The model described by Eqs.~(\ref{eq:gomp}, \ref{eq:pivot}, \ref{eq:tilt}) exhibits the expected parameter dependencies, e.g. an increase in structure formation will lead to an earlier reionization timeline. We visualize these trends explicitly in Appendix \ref{app:trends}. Similarly to other machine learning algorithms, our symbolic regression results depend on the training data and carry the risk of overfitting to it. In Appendix \ref{app:overfit}, we compute an alternative mapping using only half of our data to assess this risk. 

Armed with these rescaling, and their dependence on the cosmological parameters, we can now use our robust Gompertzian reionization model for CMB analyses.  

\section{Data \& Methodology}
\label{sec:meth}
To re-analyze state-of-the-art cosmological data with our Robust Gompertzian reionization scenario we require a Boltzmann solver capable of computing the theoretical CMB angular power spectra, the cosmological likelihoods for both CMB and external datasets, and a sampler to obtain the corresponding cosmological constraints. In this section, we introduce our nomenclature and describe our choices for each of these necessary ingredients.

\subsection{CMB data}
\label{ssec:cmb}
Throughout this work we make extensive use of a combination of CMB temperature, polarization, and lensing measurements. In what follows, we describe the components of our CMB data set, as well as the specific combination used throughout this work. 

\begin{itemize}
% \item \textbf{SPT-3G 2018 TTTEEE:} SPT-3G observations over a $\approx 1500$ deg$^2$ field of CMB temperature, polarization, and their cross-correlation. The measurements cover angular scales between $750 \leq \ell \leq 3000$ \cite{2023PhRvD.108b3510B}. We use candl\footnote{\url{https://github.com/Lbalkenhol/candl}} \cite{2024arXiv240113433B} to interface this likelihood with cobaya. 
% \item \textbf{ACT DR4 TTTEEE:} ACT DR4 $C_{\ell}^{\rm TT}$, $C_{\ell}^{\rm TE}$, and $C_{\ell}^{\rm EE}$ measurements. The maps cover approximately 17000 deg$^2$ and the power spectra focuses on small scales $\ell > 600$ \cite{2020JCAP...12..045C,2020JCAP...12..047A}. We use candl to pass this likelihood to cobaya. 
\item \textbf{\boldmath Planck low-$\ell$:} Planck PR3 CMB measurements of $C_{\ell}^{\rm TT}$ and $C_{\ell}^{EE}$ at large scales ($2 \leq \ell \leq 30$). The temperature measurements are obtained by the \texttt{Commander} Planck likelihood \cite{2020A&A...641A...5P} while the polarization one is given by \texttt{Sroll2} \citep{2019A&A...629A..38D}. 
\item \textbf{Planck PR4:} The Planck PR4 TTTEEE, i.e. CMB temperature, polarization, and cross-correlation measurements at high-$\ell$ using the NPIPE pipeline \cite{2020A&A...643A..42P} over $\approx 80$ \% of the sky. The data is based on the \texttt{CamSpec} likelihood \cite{2019arXiv191000483E,2022MNRAS.517.4620R} and covers $30 \leq \ell \leq 2500$. 
\item \textbf{ACT DR6:} The Atacama Cosmology telescope (ACT) Data Release 6 (DR6) CMB TTTEEE measurements \citep{2025arXiv250314452L}. The DR6 maps cover almost half the sky, but only roughly half of that is used after masking the galactic and extragalactic components.
\end{itemize}
Following \cite{2025arXiv250314452L}, we use a combination of Planck PR4 and ACT DR6 data for high-$\ell$ TTTEEE. We further complement this state-of-the-art CMB dataset with the Planck low-$\ell$ TT and EE measurements. We refer to this combination simply as `CMB'.   

For CMB lensing \cite{2003PhRvD..68h3002H}, we have the following components.
\begin{itemize}
\item \textbf{ACT DR6 lensing:} ACT DR6 CMB lensing measurements from redshifts $z \sim 0.5-5$ over 9400 deg$^2$ of the sky \citep{2024ApJ...962..112Q,2024ApJ...962..113M,2023arXiv230405196M,2023arXiv230905659F}. Specifically, we use the \emph{baseline} variant.
\item \textbf{Planck PR4 lensing:} Reconstruction of CMB lensing with $100 \leq \ell \leq 2048$ \cite{2022JCAP...09..039C}. The Planck NPIPE lensing map leads to an increase of signal-to-noise ratio of around 20\% compared to Planck PR3 \cite{2020A&A...641A...8P}. This NPIPE lensing map covers 65\% of the total sky area.
\end{itemize}
Following \cite{2023arXiv230405196M, 2024ApJ...962..112Q,2023arXiv231206482G}, we utilize the combined NPIPE and ACT DR6 lensing measurements. This combination is justified since NPIPE and ACT DR6 observations have minimal overlap in the sky and probe primarily different angular scales. Hence, they provide nearly independent lensing measurements that are consistent, which allows for the combination of the two observations at the likelihood level. Throughout this work we refer to this CMB lensing combination as `lensing'.

\subsection{\boldmath Low-\texorpdfstring{$z$}{z} data}
\label{ssec:lowz}
We complement the statistical power of high-redshift CMB data with low-redshift counterparts. We consider two sources of cosmological information at low redshifts: First, the Baryon Acoustic Oscillation (BAO) imprinted on large-scale correlations of cosmological tracers. Second, Type Ia supernovae, which function as standard candles to calibrate distances on cosmological scales. While BAO probes generally provide stronger statistical power within $\Lambda$CDM, Type Ia supernovae offer valuable constraints on low-redshift extensions — such as evolving dark energy \cite{2020A&A...641A...6P}. They help anchor the background cosmology at lower redshifts where conventional galaxy surveys lack the volume for precise BAO measurements. Given the recent interest in deviations from $\Lambda$CDM \cite{2020A&A...641A...6P,2024JCAP...09..053W,2024arXiv241107970Z,2025JCAP...02..021A,2025arXiv250314738D,2025arXiv250314454C,2025arXiv250504275Q}, we also include supernovae data to facilitate comparison with other efforts to investigate dark energy models. 
\begin{itemize}
    \item \textbf{BAO:} The Dark Energy Spectroscopic Instrument (DESI) \cite{2022AJ....164..207D} BAO measurements \citep{2025arXiv250314739D,2025arXiv250314738D} from the second data release. It includes BAO information from bright galaxies, luminous red galaxies, emission line galaxies, quasars, and the Lyman-$\alpha$ forest. The measurements cover the range $0.1 \lesssim z \lesssim 4$. %The Dark Energy Spectroscopic Instrument (DESI) \cite{2022AJ....164..207D} BAO measurements from the year 1 data release \cite{2025arXiv250314745D}. It includes BAO information from bright galaxies, luminous red galaxies, emission line galaxies, quasars \cite{2024arXiv240403000D}, and from the Lyman-$\alpha$ forest \cite{2024arXiv240403001D} encompassing over 6 million extragalactic objects between $0.1 \leq z \leq 4.2$ \cite{2025JCAP...02..021A}.
    \item \textbf{snD:} A homogeneously selected sample of 1635 photometrically-classified SN Ia from the Dark Energy Survey as part of their Year 5 data release \cite{2024ApJ...973L..14D}. It covers the redshift range $0.1 < z < 1.3$ and is complemented by 194 low-redshift SN Ia (in common with the Pantheon+ compilation). 
    \item \textbf{snP:} The Pantheon+ compilation of 1550 Type Ia supernovae in the redshift range $0.001 \leq z \leq 2.26$ \cite{2022ApJ...938..113S}. We rely on the likelihood from ref. \cite{2022ApJ...938..110B}. 
    \item \textbf{snU:} The Union3 compilation of 2087 SN Ia \cite{2023arXiv231112098R}. 
\end{itemize}
We do not combine the SN Ia samples because they are not independent as the different SN Ia compilations share a significant fraction of supernovae and have distinct analyses.

\subsection{Boltzmann solver}
\label{ssec:classy}
We implement our Gompertzian reionization timeline within the thermodynamics module of {\sc CLASS}\footnote{Available at \url{http://class-code.net}, while our modification of {\sc CLASS} with Gompertz-like reionization is available at \url{https://github.com/paulomontero/class_gomp.git}.} \cite{2011JCAP...07..034B}. When computing angular power spectra, our Gompertzian implementation of {\sc CLASS} uses the cosmological parameters to determine the pivot and tilt of the rescaling via Eqs.~(\ref{eq:pivot}, \ref{eq:tilt}), which are then used to compute the redshift evolution of the neutral hydrogen fraction following Eq.~(\ref{eq:gomp}). Thus, this framework allows {\sc CLASS} to calculate the optical depth to reionization as a function of cosmology (and reionization astrophysics).

We consider four distinct reionization models throughout this work:
\begin{itemize}
    \item \textbf{Rgomp:} Our Robust Gompertzian reionization model introduced in \S\ref{sec:gomp}. This model leverages the apparent universality of the Gompertzian reionization profile, which naturally fits astrophysical constraints on the reionization timeline. Furthermore, it uses symbolic regression to establish a mapping between reionization and cosmological parameters. The model includes three nuisance astrophysical ($\zetaUV$, $\Tvir$, and $\LX$) parameters that are marginalized over to derive cosmological constraints. 
    \item \textbf{Cgomp:} The `Classic' Gompertzian reionization model originally introduced in \cite{2024arXiv240513680M}. This model is a simplified variant of the Rgomp model, allowing only a single nuisance astrophysical parameter, the ionization efficiency $\zetaUV$. We include this model in \S\ref{sec:val}, primarily for its value as a baseline, effectively allowing us to quantify the dangers of oversimplifying the modeling of reionization.
    \item \textbf{Gomp:} This model only uses the Gompertzian shape for $x_\HI$, Eq.~(\ref{eq:gomp}), disregarding the cosmological dependence of $\ln \alpha$ and $\beta$ in Eq.~(\ref{eq:a_rescale}), and instead treating them directly as independent parameters.
    \item $\boldsymbol{\tanh}$: The standard treatment for reionization in most CMB analyzes, this model employs a symmetric hyperbolic tangent function to describe reionization \cite{2008PhRvD..78b3002L}. While convenient due to its simple shape, it fails to capture the expected asymmetric nature of reionization. In principle, the model has two nuisance parameters, the midpoint $z_{\rm reio}$ and duration $\Delta z_{\rm reio}$ of reionization, but most modern applications fix $\Delta z_{\rm reio} = 0.5$ and treat $z_{\rm reio}$ as either a sampled or derived\footnote{When sampling over the optical depth, Boltzmann solvers use bisection to determine the corresponding $z_{\rm reio}$, assuming a fixed $\Delta z_{\rm reio}$.} parameter. We also adopt this conventional strategy of sampling over $z_{\rm reio}$ when using the tanh model. 
\end{itemize}

All evaluations of {\sc CLASS} are done with the high-accuracy settings described in \cite{2025arXiv250314454C}. Furthermore, all models include a contribution from massive neutrinos with a fixed sum of the neutrino masses of $\sum m_\nu = 0.06$ eV, unless otherwise stated.

\subsection{Reionization data}
\label{ssec:astro}
The Gompertzian timeline of reionization allows us to supplement the cosmological data with astrophysical constraints on the evolution of the neutral hydrogen fraction. Note that this feat cannot be reproduced by the hyperbolic tangent model of reionization. We explicitly showcase this failure in Appendix \ref{app:tanh}.

We use the quasar damping wing [DW] constraints presented in \cite{2022MNRAS.512.5390G,2024MNRAS.530.3208G,2024A&A...688L..26S,2024ApJ...969..162D}, which were also used in \cite{2024arXiv240513680M}. It is important to highlight that other astrophysical constraints on the timeline of reionization exist, such as Lyman-$\alpha$ damping wings in high-$z$ galaxies\footnote{See \cite{2025arXiv250113899H} for a description of the limitations of galaxy damping wings as mechanisms to constrain the neutral hydrogen fraction.} or Ly$\alpha$ emitter luminosity function\footnote{Although note that the inference of the mean $x_\HI$ from this method is susceptible to cosmic variance \cite{2023ApJ...953...29B}.}; however, we opt to be conservative and restrict our analysis to consider only quasar damping wings due to their robustness compared to constraints based on high redshift galaxies. 

We will also compare our results with a reionization prescription that consists of only using the Gompertzian shape for the timeline of reionization without establishing the connection between cosmology and reionization (Gomp model). We include an additional reionization dataset to further restrict late reionization models since the tilt of the rescale -- Eq.~(\ref{eq:a_rescale}) -- can lead to extended unphysical late-ending reionization models. Hence, for analyses involving the Gomp model, we include the Lyman-$\alpha$ and Lyman-$\beta$ dark pixel [DP] upper-limit constraints on the end stages of reionization obtained in \cite{2015MNRAS.447..499M}\footnote{The inclusion of the DP constraints on $x_\HI(z)$ only marginally affect the cosmological constraints obtained with the Rgomp model. Figure \ref{fig:tanh_fail} already indicates that the best-fits of the Rgomp and Cgomp models are roughly compliant with these constraints even when not directly including them in the analysis.}.  

The usage of only the DW likelihood for the Rgomp model is a conservative choice regarding astrophysical constraints on the timeline of reionization. This choice is conservative in the sense that the error bars are not too restrictive. In principle, a very optimistic approach would be to use most available $x_\HI$ constraints, even if they may be in tension with each other, as was done in \cite{2025arXiv250821069E}. Likewise, another optimistic approach would be to also utilize the Lyman-$\alpha$ forest effective optical depth to achieve powerful constraints on the epoch of reionization \cite{2025PASA...42...49Q}; however, this disregards the forest sensitivity to cosmology since it assumes a fixed cosmology.

\subsection{Sampler}
\label{ssec:cobaya}
For cosmological inference, we use {\sc cobaya}\footnote{\url{https://cobaya.readthedocs.io/en/latest/}} \cite{2021JCAP...05..057T}, which incorporates several of the likelihoods described in \S \ref{ssec:cmb} and \S\ref{ssec:lowz}. 
We use the Metropolis-Hastings Monte Carlo Markov Chain (MCMC) sampler \cite{2002PhRvD..66j3511L,2013PhRvD..87j3529L} implemented within {\sc cobaya} to sample the posterior distribution. Furthermore, we introduce an external likelihood to incorporate astrophysical constraints on the reionization history from quasar damping wing observations, as discussed in \S\ref{ssec:astro}. The convergence of the chains is assessed using the Gelman-Rubin statistic \cite{Gelman1992}, requiring $R - 1 < 0.01$ as the stopping criterion.

Given the parameter dependence of Eqs.~(\ref{eq:pivot}, \ref{eq:tilt}), which arises from the cosmological parametrization in {\sc 21cmFAST}, and the ability of our Gomptertz-like reionization model to self-consistently compute the optical depth as a function of cosmological parameters, we sample over the set $\boldsymbol{\theta_{\rm cosmo}} = \{\sigma_8, n_{\rm s}, H_0, \omega_{\rm b}, \omega_{\rm cdm} \}$, supplemented by astrophysical parameters. For the Cgomp model, we include $\boldsymbol{\theta_{\rm astro}} = \{\zetaUV\}$, while for the Rgomp model, we extend this to $\boldsymbol{\theta_{\rm astro}} = \{\zetaUV, \Tvir, \LX \}$. In contrast, for the conventional $\tanh$ reionization framework, we instead sample the parameter set  $\boldsymbol{\theta} = \{ \ln (10^{10} \As), \ns, H_0, \omega_{\rm b}, \omega_{\rm cdm}, z_\re \}$ while the Gomp model uses $\boldsymbol{\theta} = \{ \ln (10^{10} \As), \ns, H_0, \omega_{\rm b}, \omega_{\rm cdm}, \ln \alpha, \frac{3}{2\beta} \}$\footnote{We follow this sampling choice since it is effective at handling the degeneracy between pivot and tilt.}.

\section{Comparison of reionization models in CMB analyses}
\label{sec:val}
\begin{figure}
    \centering
    \includegraphics[width=\linewidth]{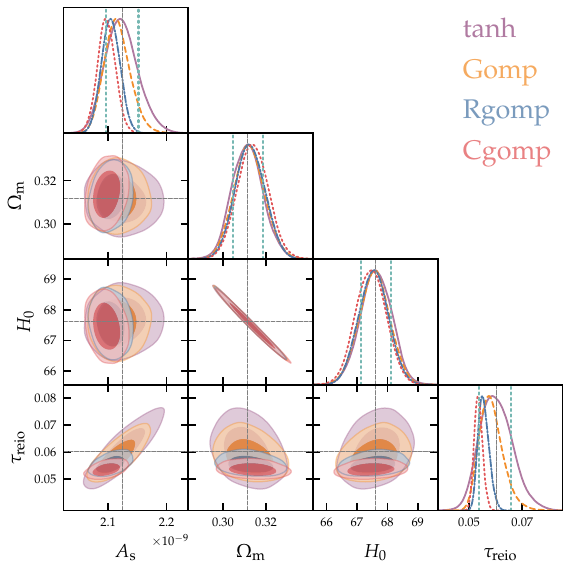}
    \caption{Representative parameter constraints based on the Cgomp \cite{2024arXiv240513680M} (red contours), Rgomp \S\ref{sec:gomp} (blue contours), Gomp (orange contours), and $\tanh$ (purple contours) reionization models. In addition to CMB data, the Gompertzian reionization models also include astrophysical constraints [DW] \S\ref{ssec:astro} -- and [DP] for the Gomp model -- on the timeline of reionization. The parameters chosen are representative ones to highlight some of the largest deviations present in Table \ref{tab:vali}. The green dashed lines in the posteriors indicate the $1\sigma$ range obtained by the ACT collaboration (`P-ACT' column in Table 5 of \cite{2025arXiv250314452L}). The inclusion of the  Gompertzian reionization timeline allows for simultaneous good fits to CMB and astrophysical observations which boosts the constraining power, while the inclusion of the missing link between $\tau$ and cosmology further improves the constraining power on cosmological parameters.}
    \label{fig:validation}
\end{figure}

To evaluate the performance of the Rgomp model with respect to other reionization models used in CMB analyses, we consider only CMB measurements (\S\ref{ssec:cmb}) and use {\sc cobaya} to perform Bayesian inference. Furthermore, we consider only $\Lambda$CDM in this section for clarity. We tabulate our results in Table \ref{tab:vali} and highlight the main inference differences in Figure \ref{fig:validation}. 

\begin{table}
    \centering
    \caption{Validation of the Rgomp reionization model. Constraints in the cosmological -- and astrophysical (nuisance) -- parameters obtained using CMB data for the conventional approach ($\tanh$ reionization), the Gompertzian shape only (Gomp model), the Cgomp reionization model introduced in \cite{2024arXiv240513680M}, and the Rgomp model (\S\ref{sec:gomp}). The Gompertzian reionization models also use astrophysical data [DW] -- and [DP] data for the Gomp model. We remind the reader that the $\tanh$ model does not include observational constraints on the reionization timeline since it does a poor job of fitting them -- see Figure 4 of \cite{2024arXiv240513680M} and Appendix \ref{app:tanh}. The numbers in parentheses give the marginalized $1\sigma$ uncertainty in the last two significant digits.}
    
    \begin{tabular}{c|cccc}
    \toprule
        Parameters & Cgomp & Rgomp & Gomp & $\tanh$  \\
        \midrule
        $10^9 \As$ & 2.097(15) & 2.105(16) & $2.116^{(19)}_{(24)}$ & $2.123^{(25)}_{(30)}$ \\ 
        $\ns$ & 0.9695(37) & 0.9697(36) & 0.9700(36) & 0.9705(37) \\
        $\Omegac h^2$ & 0.1197(12) & 0.1195(12) & 0.1194(12) & 0.1193(12)  \\
        $\Omegab h^2$ & $0.02247(11)$ & 0.02248(11) & 0.02248(11) & $0.02248(11)$ \\ 
        $\Omegam$ & 0.3139(71) & 0.3126(69) & 0.3123(71) & 0.3114(71) \\
        $H_0$ & 67.46(50) & 67.55(49) & 67.57(50) & 67.63(51) \\
        $100\theta_{\rm X}$ & 1.04161(26) & 1.04163(25) & 1.04164(25) & 1.04164(25) \\ 
        $\sigma_8$ & 0.8103(51) & 0.8110(50) & 0.8131(57) & $0.8141^{(59)}_{(66)}$ \\
        $S_8$ & 0.829(14) & 0.828(13) & 0.830(14) & 0.829(14) \\
        \hline
        \multicolumn{5}{c}{Astrophysical parameters}\\
        \hline
        $\tau_{\reio}$ & $0.0535_{(14)}^{(17)}$ & $0.0555^{(17)}_{(23)}$ & $0.0585^{(34)}_{(51)}$ & $0.0602^{(55)}_{(67)}$ \\
        $\zetaUV$ & $26.8^{(2.0)}_{(2.6)}$ & $< 24.4$ & \\
        $\Tvir$ & & $4.56^{(12)}_{(6)}$ & \\
        $\LX$ & & $< 40.2$ & \\
        $\ln \alpha$ & & & $-2.064^{(25)}_{(19)}$ & \\
        $\beta$ & & & $6.97^{(0.91)}_{(1.5)}$ & \\
    \bottomrule
    \end{tabular}
    \label{tab:vali}
\end{table}

Our CMB likelihood corresponds to the data used in the `P-ACT' analysis of \cite{2025arXiv250314452L}, thus it is reassuring that our reference $\tanh$ model successfully recovers the P-ACT cosmological constraints, as shown in Figure \ref{fig:validation}. Similarly, the Gompertzian reionization models mostly align with these constraints, except for parameters related or degenerate with the impact of reionization. In particular, Gompertzian models prefer a smaller amplitude of the primordial power spectrum ($\As$), with an accompanying decrement in the derived optical depth. Additionally, they preferentially select slightly larger values of the physical cold dark matter density ($\Omegac h^2$) and of the total matter density $\Omegam$, which likely drives the preference for a slightly smaller Hubble constant ($H_0$). Figure \ref{fig:validation} further emphasizes the importance of reionization modeling, since we see significant shifts for lower values of $\tau_\reio$ that can lead up to a $\sim1\sigma$ shift in terms of the P-ACT official results ($\tau_\reio = 0.0603^{+0.0055}_{-0.0065}$) for the Cgomp model. Coincidentally, our Rgomp results align nicely with Planck's $\tau_\reio = 0.051\pm 0.06$ \cite{2020A&A...643A..42P} and the CLASS telescope recent result of $\tau_\reio = 0.053^{+0.018}_{-0.019}$ \cite{2025ApJ...986..111L}, although these constraints rely on the $\tanh$ prescription and disregard the link between cosmology and reionization. Based on our findings here, and consistent with \cite{2024arXiv240513680M,2025arXiv250821069E}, constraints obtained with $\tanh$ shift to smaller values of $\tau_\reio$ when applying a physically-motivated reionization model due to the extended contribution to the optical depth integral from the asymmetric shape and the impact on the degeneracies of $\tau_\reio$ with cosmological parameters \cite{2020JCAP...09..005P}. Note that this trend for physically-motivated models remains even in the absence of CMB data, as was demonstrated in \cite{2025arXiv250821069E} by using astrophysical, BAO, and big bang nucleosynthesis data.

Unsurprisingly, all parameters exhibit the same trend, either Cgomp or $\tanh$ infer the largest value for a given parameter with Rgomp and Gomp having a value well within 1$\sigma$. Moreover, Figure \ref{fig:validation} showcase the benefits of adding reionization data (Gomp vs $\tanh$) and the gains in constraining power by using the connection between reionization and cosmology (e.g., Rgomp vs $\tanh$).

\begin{table}
    \caption{Goodness of fit for the CMB likelihood of the models shown in Table \ref{tab:vali}. We also include the $\chi^2$ for the DW likelihood for the Gompertzian models and the dark pixel data for the Gomp model. The main improvement is evidently in fitting the low-$\ell$ polarization data since its behaviour is strongly constrained by $\tau_\reio$. Appendix \ref{app:tanh} justifies the absence of a $\tanh$ model using both CMB and reionization data.}
    \centering
    \begin{tabular}{l|cccc}
    \toprule
        Likelihood & Cgomp & Rgomp & Gomp & $\tanh$ \\
        \hline
        \multicolumn{5}{c}{\textbf{CMB}}\\
        \hline
        $\chi^2_{\rm Planck \ PR4 \ cut}$ & 224.2 & 223.9 & 223.9 & 223.3 \\
        $\chi^2_{\rm ACT \ DR6}$ & 161.5 & 161.4 & 161.0 & 161.0 \\
        $\chi^2_{\mathrm{TT \ low}\ell}$ & 22.24 & 22.22 & 22.18 & 22.54 \\
        $\chi^2_{\mathrm{EE \ low}\ell}$ & 390.26 & 390.07 & 390.6 & 391.4 \\
        $\chi^2_{\rm CMB \ total}$ & 798.2 & 797.6 & 797.7 & 798.4 \\
        \hline
        \multicolumn{5}{c}{\textbf{Reionization}}\\
        \hline
        $\chi^2_{\rm DW}$ & 3.6 & 3.1 & 4.6 & \\
        $\chi^2_{\rm DP}$ & & & 11.3 & \\
    \bottomrule
    \end{tabular}
    \label{tab:vali_chi2}
\end{table}

Performance-wise the Rgomp and Cgomp reionization models significantly reduce the error budget in $\tau_\reio$ and $\As$ by roughly a factor of 2 and 4 compared to $\tanh$, respectively. Furthermore, in terms of goodness of fit, the Gompertzian reionization models do modestly outperform the tanh model for our selected CMB likelihood as shown in Table \ref{tab:vali_chi2}. Interestingly, the inclusion of more astrophysical parameters does not degrade the performance of the Bayesian inference of cosmological parameters significantly, although it does change the center values slightly.  In contrast, the astrophysical parameters do get altered. For instance, the constraint in $\zetaUV$ from Cgomp becomes an upper limit once Rgomp, a more robust model of reionization, is taken into account. Nonetheless, the Rgomp model is still able to shed light into the astrophysics of reionization with a compelling constraint $\log(\Tvir) = 4.56^{+0.12}_{-0.06}$, a strong upper limit in $\zetaUV$, and an informative $\LX$ upper limit, which could influence model selection for 21 cm cosmology -- see, e.g., \cite{2022ApJ...924...51A,2025arXiv250500373G,2025arXiv250710533M} for the recent status of 21 cm upper limits and their implications for the intergalactic medium. Note that there is a symbiotic relationship between our methodology and 21 cm cosmology, as future direct measurements of the epoch of reionization will constrain the astrophysical parameters introduced in \S\ref{sec:gomp} while simultaneously providing measurements on $x_\HI(z)$. 

Assuming X-ray heating is sourced by hot massive X-ray binaries (HMXBs), we note that the upper limit for the X-ray heating, $\LX < 40.2$, indicates a preference for smaller $\LX$ than expected from theoretical arguments for low-metalicity HMXBs $\langle \LX \rangle = 40.5$ \cite{2013ApJ...764...41F}. However, this upper bound is well-aligned with the observed average from local HMXBs $\langle \LX \rangle \approx 39.5 $ \cite{2012MNRAS.419.2095M}. Nevertheless, this finding is in slight tension with the HERA Phase I limits \cite{2022ApJ...924...51A}, which disfavor $\LX$ values from local, metal-rich galaxies at more than $1\sigma$. This discrepancy is likely caused by both our methodology's poor sensitivity to $\LX$ and by the usage of a distinct CMB data set\footnote{The HERA collaboration used Planck PR3.}. Besides, the HERA collaboration's result relies not only on the upper limits of the 21 cm power spectrum but also on the use of the complementary data -- UV luminosity function, EoR history, global 21 cm signal, and CMB optical depth data -- to reduce the available parameter space. 

We underscore that the upper limit on $\zetaUV$ indicates a strong preference for late reionization, in agreement with recent expectations from high redshift Lyman-$\alpha$ forest findings \cite{2020MNRAS.491.1736K,2020MNRAS.494.3080N,2021ApJ...917L..37C,2022ApJ...932...76Z,2022MNRAS.514...55B,2025ApJ...981L..27G}. Furthermore, we emphasize that our results naturally depend on the data considered. This is why we only consider the quasar damping wings constraints on the timeline of reionization for the Cgomp and Rgomp models since the error bars are not too restrictive, compared, for example, to constraints in $x_\HI(z)$ from UV/Ly$\alpha$ luminosity function \cite{2021ApJ...919..120M}. The \emph{conservative} nature of the quasar damping wings is a consequence of being direct measurements of IGM absorption, which directly correlate with the ionization state of the IGM. In contrast, other constraints may rely in a more indirect mapping between neutral hydrogen history and the observables, such as constraints based on luminosity functions that may be impacted by observational completeness and map the evolution of luminosity function, disentangled from the evolution due to hierarchical structure formation, to the neutral hydrogen history.

To finish this section, we would like to emphasize that our optical depth results for Rgomp and Cgomp models are essentially on par with the anticipated performance of \emph{next-generation} CMB satellites, such as LiteBIRD \citep{2023PTEP.2023d2F01L} and CORE \citep{2018JCAP...04..017D}, which both aim to reach the cosmic variance limit: $\sigma(\tau_\reio) = 0.002$. Note that this limit assumes the $\tanh$ reionization prescription, thus disregarding the information present in the coupling between cosmological parameters and reionization astrophysics.

\section{Into the Gompverse: Joint analysis of CMB, CMB lensing, BAO, and astrophysical data}
\label{sec:results}
Having established that our robust Gompertzian model performs as intended, we now focus on investigating its impact for different datasets and cosmological models. Naturally, the ability of Rgomp to tightly constrain the optical depth encourages exploration of its interface with the sum of the neutrino masses, which is highly degenerated with $\tau_\reio$. Besides, due to the recent interest in dynamical dark energy, we also explore the coupling between our Rgomp model and the background evolution of the Universe. Recent analyses combining CMB + lensing + BAO data have reported a preference for $w_0w_a$CDM \cite{2025arXiv250314738D}, which in turn might indicate a tendency toward negative values of the sum of the neutrino masses \cite{2025PhRvD.111h3507G}. Here, we investigate to what extend these preferences, if present, are driven or influenced by the reliance on the $\tanh$ reionization model.

In Table \ref{tab:results}, we tabulate the inferred cosmological and astrophysical parameters, as well as the corresponding $\chi^2$ for the combination of CMB + lensing + BAO + DW obtained using Rgomp. We include four different \emph{cosmologies}: $\Lambda$CDM, $\Lambda$CDM with \emph{free} sum of the neutrino masses, evolving dark energy following the Chevallier-Polarski-Linder parameterization \cite{2001IJMPD..10..213C,2003PhRvL..90i1301L}, and evolving dark energy with free sum of the neutrino masses. Note that our Rgomp model -- or more precisely our simulation suite -- does not account for the impact of massive neutrinos; nonetheless, we justify the usage of this model for our free $\sum m_\nu$ models by the small impact they have on the reionization history (see Figure \ref{fig:Mnu_xHI} and Appendix \ref{app:Mnu}). 

Similarly, our simulation suite assumes $\Lambda$CDM as the background to compute the $x_\HI$ samples. We justify this decision by the fact that reionization is completed well into the matter domination era. Appendix \ref{app:w0wa} further reinforces this point by establishing the equality redshift between evolving dark energy and matter. 

Since we are now comparing the same dataset `CMB + lensing + BAO + DW' across different models, we will complement the goodness of fit of the different likelihoods with the Bayesian evidence\footnote{We use MCEvidence \cite{2017arXiv170403472H} \url{https://github.com/yabebalFantaye/MCEvidence}.}. Naturally, we warn that interpretations of the Bayes' factor using the Jeffreys' scale are intrinsically linked to our choice of priors, so readers should keep this in mind for the relevant statements in this section. For the purpose of computing Bayesian evidence ratios, we run a $\tanh$ model with $\Lambda$CDM, obtaining a Bayesian evidence of $\ln \mathcal{Z}^{\Lambda \mathrm{CDM}}_{\rm tanh} \approx -431.14$. In contrast, the analog Rgomp analysis yields $\ln \mathcal{Z}_{\rm Rgomp}^{\Lambda \mathrm{CDM}} \approx -423.57$, corresponding to a Bayesian evidence ratio of $\Delta \ln \mathcal{Z} = \ln \mathcal{Z}^{\Lambda \mathrm{CDM}}_{\rm Rgomp} - \ln \mathcal{Z}^{\Lambda \mathrm{CDM}}_{\rm tanh} = 7.57$, i.e. decisive evidence in favor of our Rgomp model over the $\tanh$ parametrization according to Jeffreys' scale \cite{jeffreys1998theory}. 

\begin{table}[]
    \centering
    \caption{Parameter constraints and $\chi^2$ for the Rgomp model with CMB + lensing + BAO + DW considering $\Lambda$CDM, \emph{free} massive neutrinos, evolving dark energy, and free sum of the neutrino masses with evolving dark energy. We also include the individual $\chi^2$ contributions for the different components of our complete likelihood and the Bayesian evidence $\ln \mathcal{Z}$. For reference a $\tanh$ run with the same dataset yields a Bayesian evidence of $\ln \mathcal{Z} \approx -431.14$. Furthermore, to facilitate a comparison of the non-reduced $\chi^2$ of the different likelihoods, we remind the reader that $\sum m_\nu \Lambda$CDM adds an additional parameter compared with $\Lambda$CDM, while $w_0w_a$CDM and $\sum m_\nu w_0w_a$CDM add two and three parameters, respectively.  The numbers in parentheses give the marginalized $1\sigma$ uncertainty in the corresponding last significant digits. $\dag$ indicates that the parameter is fixed to the given value.}
    \begin{tabular}{l|cccc}
    \toprule
     Parameters & $\Lambda$CDM & $\sum m_\nu\Lambda$CDM & $w_0w_a$CDM & $\sum m_\nu w_0w_a$CDM \\
     \hline
      $10^9 \As$ & 2.109(12) & 2.105(12) & 2.099(12) & 2.100(12)\\
      $\ns$ & 0.9740(30) & 0.9731(30) & 0.9704(32) & 0.9704(32) \\
      $\Omegac h^2$ & 0.11754(63) & 0.11787(64) & 0.11921(81) & 0.11915(81) \\
      $\Omegab h^2$ & $0.02256^{(10)}_{(11)}$ & 0.02250(10) & 0.02249(11) &  0.02250(10) \\
      $\Omegam$ & 0.3011(35) & 0.2992(36) & $0.347^{(21)}_{(17)}$ & $0.347^{(23)}_{(18)}$ \\
      $H_0$ & 68.37(27) & 68.57(28) & $64.1^{(1.4)}_{(2.0)}$ & $64.1^{(1.4)}_{(2.2)}$ \\
      $100\theta_\mathrm{X}$ & 1.04179(24) & 1.04174(25) & $1.04165(25)$ & 1.04166(25) \\
      $\sigma_8$ & $0.8063^{(31)}_{(28)}$ & $0.8151^{(60)}_{(34)}$ & $0.783^{(12)}_{(16)}$ & $0.781^{(16)}_{(20)}$ \\
      $S_8$ & $0.8078^{(72)}_{(66)}$ & 0.8140(74) & $0.842^{(12)}_{(11)}$ & $0.839{(12)}$ \\
      $\tau_\reio$ & $0.0564^{(19)}_{(23)}$ & $0.0561^{(17)}_{(22)}$ & $0.0552^{(18)}_{(23)}$ & $0.0553^{(18)}_{(22)}$ \\
      $\zetaUV$ & $<24.6$ & $<24.4$ & $<25.4$ & $< 25.8$ \\
      $\Tvir$ & 4.53(9) & $4.56^{(12)}_{(7)}$ & $4.53^{(12)}_{(9)}$ & $4.52^{(12)}_{(95)}$ \\
      $\LX$ & $<40.2$ & $<40.2$ & $<40.2$ & $<40.3$ \\
      $w_0$ & & & $-0.47^{(20)}_{(17)}$ & $-0.47^{(22)}_{(18)}$ \\
      $w_a$ & & & $-1.57(53)$ & $-1.60(60)$ \\
      $\sum m_\nu$ [eV] & $0.06^\dag$& $<0.0274$ & 0.06$^\dag$ & $< 0.0942$ \\
      \hline 
      \multicolumn{5}{c}{Goodness of fit}\\
      \hline
      $\chi^2_{\rm Planck \ PR4 \ cut}$ & 223.4 & 223.0 & 223.4 & 223.4 \\
      $\chi^2_{\rm ACT \ DR6}$ & 164.0 & 162.5 & 160.2 & 160.5 \\
      $\chi^2_{\mathrm{TT \ low}\ell}$ & 21.51 & 21.71 & 22.14 & 22.11 \\
      $\chi^2_{\mathrm{EE \ low}\ell}$ & 390.13 & 390.11 & 390.08 & 390.07 \\
      $\chi^2_{\rm lensing}$ & 20.6 & 20.4 & 20.15 & 20.14 \\
      $\chi^2_{\rm BAO}$ & 12.9 & 11.9 & 9.1 & 9.4 \\
      $\chi^2_{\rm DW}$ & 3.3 & 3.1 & 3.0 & 3.1 \\
      $\ln \mathcal{Z}$ & -423.57 & -425.69 & -424.62 & -429.08 \\ 
    \bottomrule
    \end{tabular}
    \label{tab:results}
\end{table}

\subsection{Impact on \texorpdfstring{$\Lambda$}{L}CDM}
\label{ssec:lcdm}
\begin{figure}
    \centering
    \includegraphics[width=0.7\linewidth]{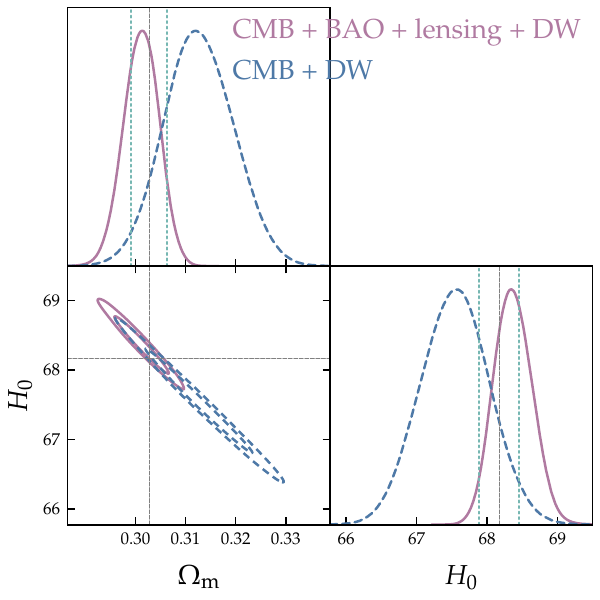}
    \caption{Impact of the inclusion of BAO and CMB lensing on the total matter density and Hubble constant. The solid purple (dashed blue) lines represent the posterior/contour for the analysis of CMB + BAO + lensing + DW (CMB + DW) with the Rgomp reionization model. The gray dotted lines indicate the central value inferred by DESI DR2 \cite{2025arXiv250314738D} [DESI + CMB] analysis, with the green lines marking the $1\sigma$ range. Note that the inclusion of BAO also yields larger $\ns$ and $\tau_\reio$ \cite{2025arXiv251205108M}.}
    \label{fig:Omegam_H0}
\end{figure}

Before comparing the different columns of Table \ref{tab:results}, we highlight that the inclusion of CMB lensing and BAO data leads to some small deviations compared to Table \ref{tab:vali}. Particularly, the supplementary data increase the baryon density, amplitude and tilt of the primordial power spectrum, and Hubble parameter, while reducing the matter density and cold dark matter density. The trends present in the expansion/density parameters are easily explainable by the introduction of BAO data since it breaks the geometric degeneracy between $H_0$ and $\Omegam$. Specifically, we obtain 
\begin{eqnarray}
\label{eq:H0}
\begin{rcases}
    \Omegam  =  0.3011 \pm 0.0035, \\
    H_0  =  68.37 \pm 0.27 \ \ \ \  \textup{[km s}^{-1} \, \textup{Mpc}^{-1} \textup{]}
\end{rcases} \ \mathrm{(CMB + BAO + lensing + DW).}
\end{eqnarray}
For reference, DESI DR2 obtained $\Omegam = 0.3027 \pm 0.0036$ and $H_0 = 68.17 \pm 0.28$ when combining DESI with CMB data \cite{2025arXiv250314738D}. Figure \ref{fig:Omegam_H0} contextualizes our findings in light of the DESI DR2 $1\sigma$ range. We infer a larger value for the Hubble constant, very slightly more aligned with the SH0ES supernovae measurements calibrated using Cepheids ($H_0 = 73.17 \pm 0.86$ km s$^{-1}$ Mpc$^{-1}$  \cite{2024ApJ...973...30B}). Likewise, our findings better align with the Chicago-Carnegie Hubble program using TRGB (Tip of the Red Giant Branch) stars for calibration \cite{2019ApJ...882...34F}. In particular, their recent measurements using solely James Webb Space Telescope \cite{2006SSRv..123..485G} data yield $H_0 = 68.81 \pm 2.22$ \cite{2025ApJ...985..203F} in good agreement with Eq.~(\ref{eq:H0}).  

The inclusion of BAO can also lead to smaller $\As$ \emph{and larger} $\tau_\reio$. Ultimately, this is a consequence of the breaking of the $\Omegam - H_0$ degeneracy. The BAO preference for smaller $\Omegam$ yields lower clustering, and hence lower $\sigma_8$, while simultaneously decreasing the $k_{\rm eq}$, the scale of equality, and also delaying the equality (between matter and radiation). The former enhances oscillations at large scales, increasing the relative strength of the first acoustic peaks \cite{1996ApJ...471..542H}, which prompts counterbalance by a decreasing $\As$ and rising $\ns$. However, larger values of $\ns$ also yield a high-$\ell$ enhancement in $C_\ell^{\rm TT}$, causing $\tau_\reio$ to rise to compensate by suppressing the high-$\ell$ TT angular power spectrum. Furthermore, the inclusion of BAO also affects the polarization angular power spectra, via changes to equality and the integrated Sachs-Wolfe effect, yielding a decrease in the low-$\ell$ $C_\ell^{\rm EE}$; however, the rise of $\tau_\reio$ can already compensate for this as the amplitude of the low-$\ell$ $C_\ell^{\rm EE} \propto \tau_\reio^2$.

Consequently, the inclusion of BAO drives $\tau_\reio$ toward larger values \cite{2025arXiv251205108M}, therefore favoring earlier reionization scenarios. This trend is consistent with the findings of \cite{2025arXiv250309971Z}, which identified a similar preference through the impact of long-lived reionization relics on the eBOSS DR16 Lyman-$\alpha$ forest BAO correlations \cite{2020ApJ...901..153D}. 

For the astrophysical parameters, we report a slight improvement in the constraints with the addition of BAO and CMB lensing data.

\subsection{Impact on \texorpdfstring{$\sum m_\nu$}{Mnu}CDM}
\label{ssec:mnu}
Massive neutrinos suppress structure formation on scales smaller than their free-streaming scale, leaving imprints in matter tracers \cite{1998PhRvL..80.5255H,2006PhR...429..307L}. However, their effects are degenerate with other parameters that regulate the amplitude of the primordial power spectrum, such as $\As$ and $\tau_\reio$. Because of the gains on $\tau_\reio$ and $\As$ obtained in \S\ref{sec:val}, we expect a significant improvement on neutrino mass constraints relative to other works that use the hyperbolic tangent reionization prescription. Massive neutrinos also impact geometry, particularly by affecting the angular scale of the sound horizon for CMB analyses -- leading to a well-known degeneracy between $H_0$ and $\sum m_\nu$ \cite{2006PhR...429..307L,2024JCAP...12..048L,2025arXiv251015835S}. 

\begin{figure}
    \centering
    \includegraphics[width=0.8\linewidth]{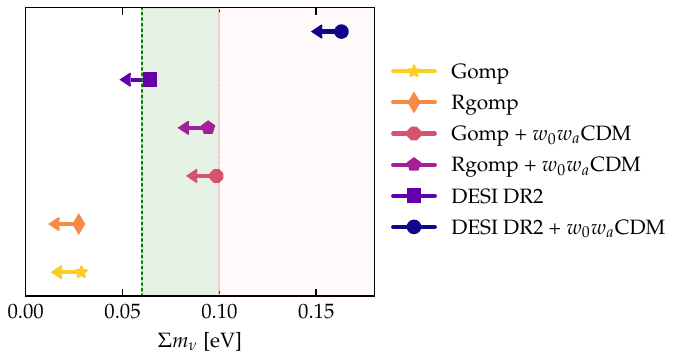}
    \caption{Constraints (95\% limits) on the sum of the neutrino masses for Rgomp using CMB + BAO + lensing + DW and for Gomp (CMB + BAO + lensing + DW + DP). We include both constraints within $\sum m_\nu \Lambda$CDM and $\sum m_\nu w_0w_a$CDM. Furthermore, we incorporate the constraints from the DESI DR2 results for DESI BAO + CMB \cite{2025arXiv250314738D} for comparison. The vertical dotted lines and shaded regions represent the minimum allowed $\sum m_\nu$ for (from left to right) the normal and inverted mass hierarchies, respectively.}
    \label{fig:Mnu}
\end{figure}

In $\Lambda$CDM, the inclusion of DESI BAO leads to a tighter constraint in $\Omegam$ while also preferring lower values (see Figure \ref{fig:Omegam_H0}). This shift, for a value near the low-end tail of the $\Omegam$ CMB posterior, will then lead to ruling out all but the smallest neutrino masses in $\sum m_\nu \Lambda$CDM (as discussed in \cite{2025arXiv250314738D}). Moreover, we obtain even tighter upper bounds with our Rgomp model, since the degeneracy between neutrino masses and optical depth is also alleviated. Consequently, instead of the DESI DR2 constraint of $\sum m_\nu < 0.0642$ \cite{2025arXiv250314738D}, we obtain a stringent constraint
\begin{equation}
    \label{eq:MnuLCDM}
    \sum m_\nu < 0.0274 \ \ \ \ (\mathrm{95\%, CMB + BAO + lens + DW)} \, .
\end{equation}
Note that Eq.~(\ref{eq:MnuLCDM}) is in tension with the normal and inverted hierarchies' bounds of the minimum neutrino masses and with a posterior that peaks in the negative mass range, thus in conflict with the smallest bound from neutrino oscillation experiments, $\sum m_\nu > 0.059 $ eV (normal ordering) \cite{2024arXiv241005380E}. However, we use the conventional physical prior of $\sum m_\nu > 0$ and do not incorporate neutrino oscillation data. Interestingly, the inclusion of neutrino oscillation data -- via NuFit-6.0 \cite{2024arXiv241005380E} -- relaxes the DESI DR2 limit from $\sum m_\nu < 0.064$ to $\sum m_\nu < 0.112$ eV \cite{2025arXiv250314738D}. Moreover, disregarding sub-degree scales on CMB temperature and polarization with an agnostic approach to the value of the sound horizon at recombination can also significantly relax the preference for negative neutrino masses \cite{2025arXiv251015835S}, while the inclusion of high-redshift galaxy data from JWST appears to tighten the neutrino mass bounds \cite{2025arXiv250910836Z}. Furthermore, ref. \cite{2025arXiv251101967C} indicated that the apparent preference for negative $\sum m_\nu$ could point to inconsistency between early-time and late-time (BAO and lensing) probes. See also \cite{2025arXiv250314744E} for a discussion about the emerging moderate tension between the lower bound implied by neutrino oscillations and cosmological constraints. Figure \ref{fig:Mnu} showcases our findings for the marginalized posterior of $\sum m_\nu$ in the context of the DESI DR2 results.

% \begin{figure}
%     \centering
%     \includegraphics[width=0.5\linewidth]{Figures/Mnu_posterior.pdf}
%     \caption{Marginalized posterior for the sum of the neutrino masses for the Rgomp  using CMB + BAO + lensing + DW. We also include constraints obtained using the Gomp model -- which contains no mapping between reionization astrophysics and cosmology -- with CMB + BAO + lensing + DW + DP for comparison. The blue dashed (red solid) line depicts the posterior obtained under $\sum m_\nu \Lambda$CDM while the yellow dashed (purple solid) line corresponds to the one obtained for $\sum m_\nu w_0 w_a$CDM for the Rgomp (Gomp) model. The vertical dotted lines and shaded regions represent the  minimum allowed $\sum m_\nu$ for (from left to right) the normal and inverted mass hierarchies, respectively. The vertical dashed-dotted lines indicate the upper bounds (95\% limits) obtained using the hyperbolic tangent reionization prescription for $\sum m_\nu \Lambda$CDM (gray) and $\sum m_\nu w_0 w_a$CDM (brown) reported in the DESI DR2 results for DESI BAO + CMB \cite{2025arXiv250314738D}.}
%     \label{fig:Mnu}
% \end{figure}

Since relatively small changes in the sum of the neutrino masses negligibly affect $x_\HI(z)$ (see Appendix \ref{app:Mnu}), we conclude, from our $\sum m_\nu \Lambda$CDM experiment, that the conventional hyperbolic tangent reionization prescription is driving the $\sum m_\nu$ limit reported by DESI to larger values that hide additional tension between DESI and the neutrino hierarchies within $\Lambda$CDM. Nevertheless, we highlight that although our findings are indicative of a worsening tension, the results have been obtained without a physical model of the neutrino masses, i.e. we do not include information on the mass squared differences (such as the value of $\Delta m_{21}^2$), which could alleviate some of the tension \cite{2025arXiv250314738D}.

Moreover, as indicated in Table \ref{tab:results}, the inclusion of massive neutrinos raises the cold dark matter contribution, which can be interpreted as an attempt to compensate for the decrease in the contribution of the massive neutrinos to the overall energy budget. We report little change in the astrophysical nuisance parameters, including the optical depth to reionization.  

Additionally, the reduction in the sum of the neutrino masses within the $\sum m_\nu \Lambda$CDM leads to an increase in $\sigma_8$, from $0.8063$ to $0.8151$. This comes as expected, since matter clustering is an integrated effect through time \cite{2025PhRvD.111h3507G}, and massive neutrinos suppress structure formation by damping the matter power spectrum. For smaller values of $\sum m_\nu$, the suppression of clustering on scales below the neutrino free-streaming length is subdued relative to a scenario with $\sum m_\nu = 0.06$ eV. Consequently, the value of $\sigma_8$ increases.  

Another consequence of the decrease in the mass of massive neutrinos, and more generally of the treatment of the sum of the neutrino masses as a free parameter, is the degradation of the $\sigma_8$ constraint. The upper limit error increases by roughly a factor of two compared to the $\Lambda$CDM  constraint. The direction of this degradation of sensitivity is indicative of the fact that $\sum m_\nu$ now spans smaller values, which enables $\sigma_8$'s posterior to span larger values. 

Regarding the goodness of fit, $\sum m_\nu\Lambda$CDM provides a slightly better fit to most likelihood components compared to $\Lambda$CDM, with the exception of the low-$\ell$ TT spectrum. However, these modest improvements are insufficient to compensate for the additional degree of freedom as the Bayesian evidence ratio,  $\Delta \ln \mathcal{Z} = \ln \mathcal{Z}^{\sum m_\nu\mathrm{CDM}} - \ln \mathcal{Z}^{\Lambda\mathrm{CDM}} = -2.12$, indicates a weak preference in favor of $\Lambda$CDM.

\subsection{Impact on \texorpdfstring{$w_0w_a$}{w0wa}CDM}
\label{ssec:w0wa}
An enhanced reionization treatment does not directly affect dynamic dark energy models; nevertheless, as demonstrated in Figure \ref{fig:validation}, it can significantly change the constraints on cosmological parameters related to the evolution of the Universe. As a result, constraints on $w_0w_a$ can shift due to the reionization model employed.

\begin{figure}
    \centering
    \includegraphics[width=\linewidth]{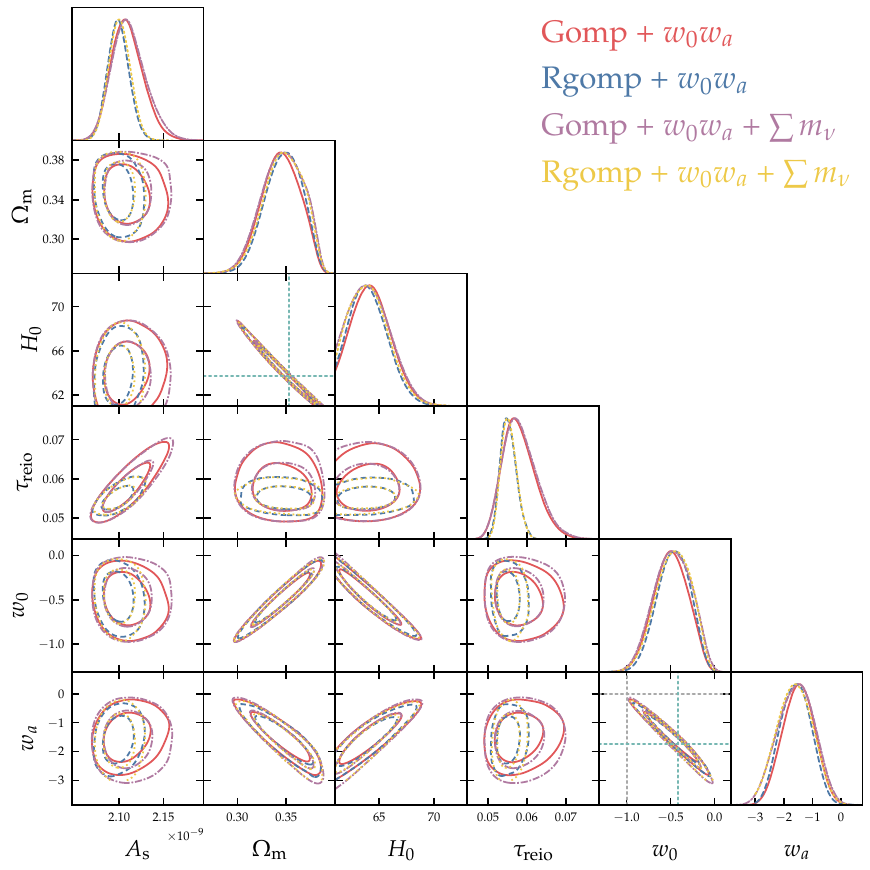}
    \caption{Triangle contour plots for representative parameters of evolving dark energy models using CMB + BAO + lensing + DW for Rgomp. For comparison, we also include the Gomp model with CMB + BAO + lensing + DW + DP. The red solid lines correspond to the inference with Gomp for $w_0w_a$CDM, while the blue dashed lines depict the contours for Rgomp with $w_0w_a$CDM. Furthermore, the yellow dotted lines (purple dashed lines) represent the constraints for Rgomp (Gomp) with $\sum m_\nu w_0w_a$CDM. The gray dotted lines correspond to the expected value of $w_0,w_a$ for $\Lambda$CDM, which is excluded by all models by more than 2$\sigma$, while the green dotted lines indicate the results from DESI DR2 \cite{2025arXiv250314738D} for $\sum m_\nu w_0w_a$ with DESI BAO + CMB.}
    \label{fig:w0wa_triangle}
\end{figure}

We remark that the apparent preference of BAO data for evolving dark energy becomes somewhat obscured when employing the Rgomp model. As shown by the goodness of fit values in Table \ref{tab:results} and the 2$\sigma$ contours in Figure \ref{fig:w0wa_triangle}, the data indicate support for $w_0w_a$CDM. However, the Bayesian evidence ratio suggests that there may be weak evidence in favor of $\Lambda$CDM, penalizing the additional degrees of freedom. Interestingly, the $w_0w_a$CDM chain also provides a better fit to high-$\ell$ ACT DR6 TTTEEE CMB, low-$\ell$ EE, CMB lensing, and quasar damping wing constraints compared to those of $\Lambda$CDM and $\sum m_\nu$CDM. The tradeoff for these improvements appears to be a slight degradation in the fit to low-$\ell$ TT data.

Regarding parameter constraints, we expectedly confirm that $H_0$ increases and $\Omegam$ decreases relative to $\Lambda$CDM. The latter contributes to an increase in clustering, which is however, compensated by the decrease in $\As$ leading to an overall lesser $\sigma_8$ in $w_0w_a$CDM. In addition, we report that $w_0w_a$CDM prefers a later reionization timeline compared to $\Lambda$CDM. Figure \ref{fig:w0wa_triangle} depicts how these trends are affected by the use of the Rgomp model compared to the Gomp model. 

Using Rgomp we report a constraint for $w_0w_a$CDM of 
\begin{eqnarray}
    \begin{rcases}
        w_0 = -0.47^{+0.20}_{-0.17} \\
        w_a = -1.57 \pm 0.53
    \end{rcases} \ \mathrm{(CMB + BAO + lensing + DW),}
    \label{eq:w0wa}
\end{eqnarray}
which indicates a preference for a \emph{weaker} evolution compared to the findings of DESI DR2's DESI + CMB ($w_0 = -0.42 \pm 0.21$ and $w_a = -1.75 \pm 0.58$) \cite{2025arXiv250314738D}. Likewise, Rgomp yields slightly stronger constraints, as is evident in the small improvement in the errors. Notably, this improvement is also present in the Gomp model ($w_0 = -0.50 \pm 0.19$ and $w_a = -1.48 \pm 0.55$). Note that our inferred values indicate that phantom crossing (e.g., \cite{2005PhRvD..71d7301H}) occurred somewhere below $z < 1$, a feature also present for DESI DR2 results. 

\subsection{Impact on $\sum m_\nu w_0w_a$CDM \& inclusion of supernovae data}
\label{ssec:Mnuw0wasn}
Even if the impact of the Rgomp model on the $w_0-w_a$ plane was limited, our physically motivated reionization model does affect the constraints on the neutrino masses significantly, as demonstrated in \S\ref{ssec:mnu}. However, allowing for a free $\sum m_\nu$, has little impact on the $w_0w_a$\footnote{Similar to what happens for the conventional hyperbolic tangent prescription \cite{2025arXiv250314738D}.} constraints from the Rgomp model
\begin{eqnarray}
    \begin{rcases}
        w_0 = -0.47^{+0.22}_{-0.18} \\
        w_a = -1.60 \pm 0.60
    \end{rcases} \ \mathrm{(CMB + BAO + lensing + DW +}  \ \sum m_\nu).
\end{eqnarray}
These small shifts compared to Eq.~(\ref{eq:w0wa}) are also accompanied by a small impact on all model parameters and corresponding $\chi^2$'s, as shown in Table \ref{tab:results}. However, according to the Jeffreys' scale ($\Delta \ln \mathcal{Z} < -5$), there is very strong evidence against the $\sum m_\nu w_0 w_a$CDM model compared to $\Lambda$CDM for our Rgomp reionization. 

Regarding the inferred sum of the neutrino mass constraint, we obtain a less stringent bound compared to Eq.~(\ref{eq:MnuLCDM})
\begin{equation}
    \label{eq:Mnuw0wa}
    \sum m_\nu < 0.0942 \ \ \ \ (\mathrm{95\%, CMB + BAO + lens + DW + } \ w_0w_a) \, ,
\end{equation}
but still lower than the limit expected from the inverted mass ordering. The relaxation of $\sum m_\nu$ constraints in dynamical dark energy models, relative to $\sum m_\nu \Lambda$CDM, arises from the additional freedom in the expansion history. This feature can be understood as a consequence of the impact of massive neutrinos on the geometry. Neutrinos modify the angular scale under which the sound horizon is observed in CMB data \cite{2025arXiv251015835S}, while BAO measurements help constrain the background evolution, which has a couple more free parameters for the Hubble rate contribution that enters the integral in the comoving angular diameter distance. 

\begin{figure}
    \centering
    \includegraphics[width=0.6\linewidth]{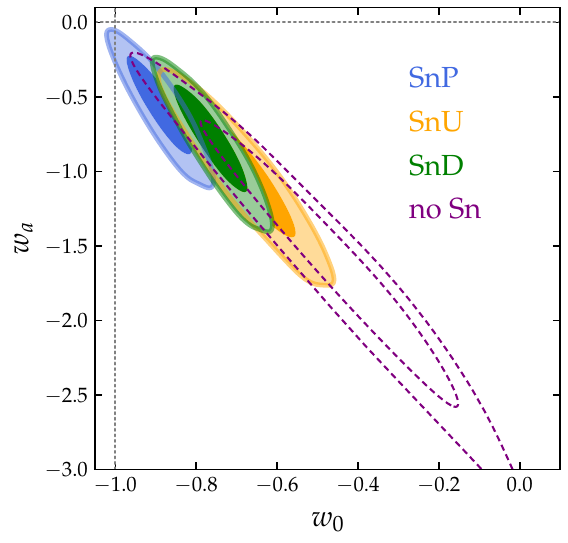}
    \caption{Contour plot of $w_0$-$w_a$ for CMB + BAO + lens + DW + different supernovae datasets using the Rgomp model with $\sum m_\nu$ as a free parameter. The blue contour corresponds to Pantheon+, while the orange and green contours represent the constraints for Union3 and DESY5, respectively. For comparison, we include the purple dashed contour that corresponds to CMB + BAO + lens + DW. The inclusion of Sn Ia data significantly tightens the constraint on dynamic dark energy while also moving the mean values closer to the $\Lambda$CDM corner. }
    \label{fig:sn}
\end{figure}

We now move to focus on the inclusion of supernovae type Ia data and their impact on $w_0w_a$ constraints employing the three snD, snP, and snU datasets described in \S\ref{ssec:lowz}. Figure \ref{fig:sn} summarizes the results in the $w_0$-$w_a$ plane. Specifically for the Union3 catalog, we obtain
\begin{equation}
    \begin{rcases}
        w_0 = -0.680^{+0.082}_{-0.092} \\
        w_a = -1.01^{+0.33}_{-0.27}
    \end{rcases} \ \mathrm{(CMB + BAO + lensing + DW + snU + }  \ \sum m_\nu),
\end{equation}
with an accompanying upper limit of $\sum m_\nu < 0.0756$ (95\% CL). 

Similarly, for DESY5, we get 
\begin{equation}
    \begin{rcases}
        w_0 = -0.760 \pm 0.060 \\
        w_a = -0.79^{+0.26}_{-0.21}
    \end{rcases} \ \mathrm{(CMB + BAO + lensing + DW + snD + }  \ \sum m_\nu),
\end{equation}
with an accompanying upper limit of $\sum m_\nu < 0.0697$ (95\% CL). Notably, the upper bound in $\sum m_\nu$ is significantly tighter to the one reported in the DESI DR2 results ($\sum m_\nu < 0.129$ eV) obtained using DESI + CMB + DESY5 \cite{2025arXiv250314738D}. Naturally, some of the difference is driven by our usage of the state-of-the-art CMB dataset (which includes the high-$\ell$ ACT DR6 multiples); nevertheless, the primary driver is the usage of the Rgomp model, which allows for the usage of astrophysical constraints on the timeline of reionization, in contrast to the standard tanh model.

For the Pantheon+ compilation, we obtain the following constraints 
\begin{equation}
    \begin{rcases}
        w_0 = -0.888 \pm 0.053 \\
        w_a = -0.57^{+0.23}_{-0.20}
    \end{rcases} \ \mathrm{(CMB + BAO + lensing + DW + snP + }  \ \sum m_\nu).
\end{equation}
Just as in the DESI DR2 results, the inclusion of the Pantheon+ supernovae data leads to the strongest upper bound on the sum of the neutrino masses ($\sum m_\nu < 0.117$ eV), which in our case becomes $\sum m_\nu < 0.0569$ at 95\% CL. Interestingly, this limit is in tension with the minimum mass allowed by neutrino flavor oscillation experiments for the normal hierarchy \cite{2024arXiv241005380E}. 

As discussed in \S\ref{ssec:mnu}, the direct inclusion of neutrino oscillation data can alleviate these tensions; however, we do not incorporate them in our analysis. Instead, we conclude that the inclusion of $w_0w_a$ does relax the bounds on the sum of the neutrino masses, but remains insufficient to fully resolve the existing neutrino tensions when a physically motivated reionization model is used. Furthermore, our findings indicate that the use of the conventional $\tanh$ reionization prescription may be obscuring the level of tension present in cosmological constraints of $\sum m_\nu$. Likewise, alternative explanations for the apparent preference of negative neutrino masses, such as a high optical depth by dropping the low-$\ell$ $C_\ell^{\rm EE}$ \cite{2025arXiv250416932S}, may be potentially biased by the $\tanh$ model.

\section{Summary}
\label{sec:sum}
Most analyses of CMB data rely on the usage of a hyperbolic tangent to model cosmic reionization. This analytical template neglects the connection between reionization and cosmology, while also proving a poor fit to astrophysical constraints on the timeline of reionization. Here we introduce Rgomp, a physically motivated Gompertzian reionization model with three nuisance parameters to account for reionization astrophysics, capable of reproducing observations of the epoch of reionization. Furthermore, the Rgomp model uses symbolic regression to exploit the connection between cosmology and reionization. We summarize our main findings as follows. 
\begin{itemize}
    \item Physically motivated models of reionization yield significantly stronger constraints on the optical depth to reionization. Compared to the conventional hyperbolic tangent prescription, we see a reduction of roughly a factor of three in the optical depth uncertainty (see Table \ref{tab:vali}), a level of precision comparable with the anticipated performance of \emph{next-generation} CMB satellites \cite{2018JCAP...04..017D,2023PTEP.2023d2F01L}. Moreover, Gompertzian reionization models tend to favor later reionization scenarios than the $\tanh$ parameterization, while simultaneously providing an improved fit to low-$\ell$ CMB temperature and polarization data. 
    \item Comparing the performance of $\tanh$ versus Rgomp models in the same dataset -- `CMB + lensing + BAO + DW' -- results in a Bayesian evidence ratio of $\Delta \ln \mathcal{Z} = -7.57$, and hence very strong evidence against the $\tanh$ prescription according to the Jeffreys' scale. 
    \item Including a more robust treatment of reionization astrophysics, through the introduction of three astrophysical nuisance parameters, leads to small shifts relative to the model with a single nuisance parameter introduced in \cite{2024arXiv240513680M} (see Figure \ref{fig:validation}). The resulting constraints and upper limits on the astrophysical parameters inferred with the Rgomp model can directly inform key parameter ranges for current and future 21 cm studies.  
    \item As shown in Figure \ref{fig:Omegam_H0}, our physically motivated reionization model for the full `CMB + BAO + lensing + DW' in $\Lambda$CDM yields a slightly larger (smaller) value of $H_0$ ($\Omegam$) than the DESI DR2 findings \cite{2025arXiv250314738D} (consistent at the 1$\sigma$ level) but still in significant tension with $H_0$ local measurements \cite{2022ApJ...934L...7R,2024ApJ...973...30B,2025ApJ...985..203F}.
    \item The inferred values of $w_0$ and $w_a$ using the Rgomp model exclude $\Lambda$CDM at more than $2\sigma$ as illustrated in Figures \ref{fig:w0wa_triangle} and \ref{fig:sn}, in agreement with other analysis that includes DESI DR2 BAO results \cite{2025arXiv250314738D,2025arXiv250314454C}. However, we report a Bayesian evidence ratio that suggests a minor preference for the $\Lambda$CDM Rgomp model rather than $w_0w_a$CDM.  
    \item By reducing the uncertainty on $\tau$, the Rgomp model leads to tight upper bounds on the sum of neutrino masses, $\sum m_\nu$, when it is varied. Within $\sum m_\nu$CDM model, we report $\sum m_{\nu} < 0.0274$ (95\% CL), which is in significant tension with neutrino oscillation experiments (Figure \ref{fig:Mnu}). Simultaneously varying $\sum m_\nu$ and $w_0w_a$ relaxes the constraint but can still lead to tensions, e.g. $\sum m_\nu < 0.0569$ for `CMB + BAO + lensing + DW + snP'.
\end{itemize}

With the increasing number and precision of astrophysical constraints on the timeline of reionization \cite{2025arXiv250414746K,2025arXiv251025829D,2025arXiv250805739C,2025arXiv250404683U,2025ApJS..278...33K}, the dawn of 21 cm power spectrum measurements from the epoch of reionization \cite{2022ApJ...924...51A, 2020PASA...37....7S,2025A&A...699A.109G,2025arXiv250804164T,2025arXiv251121289T}, and the anticipated performance of the next CMB experiments \cite{2018JCAP...04..017D,2023PTEP.2023d2F01L}, it is an opportune moment to revisit our conventional treatment of reionization for CMB analyses and to cultivate the ability to self-consistently use the combined power of these datasets. 

Based on our current understanding of the physics that govern cosmic reionization, reionization must have been an asymmetric process where its initial stages showcase a slower pace compared to its later stages. Consequently, symmetric parametrizations such as the hyperbolic tangent are inadequate for jointly fitting CMB and reionization observables. In this work, we introduce a robust and interpretable Gompertzian reionization model, characterized by three astrophysical nuisance parameters, that accurately captures this asymmetry and excels at joint fits of CMB and reionization data.

\appendix
\section{Cosmological and astrophysical trends of the mapping between reionization and cosmology}
\label{app:trends}
\begin{figure}
    \centering
    \includegraphics[width=\linewidth]{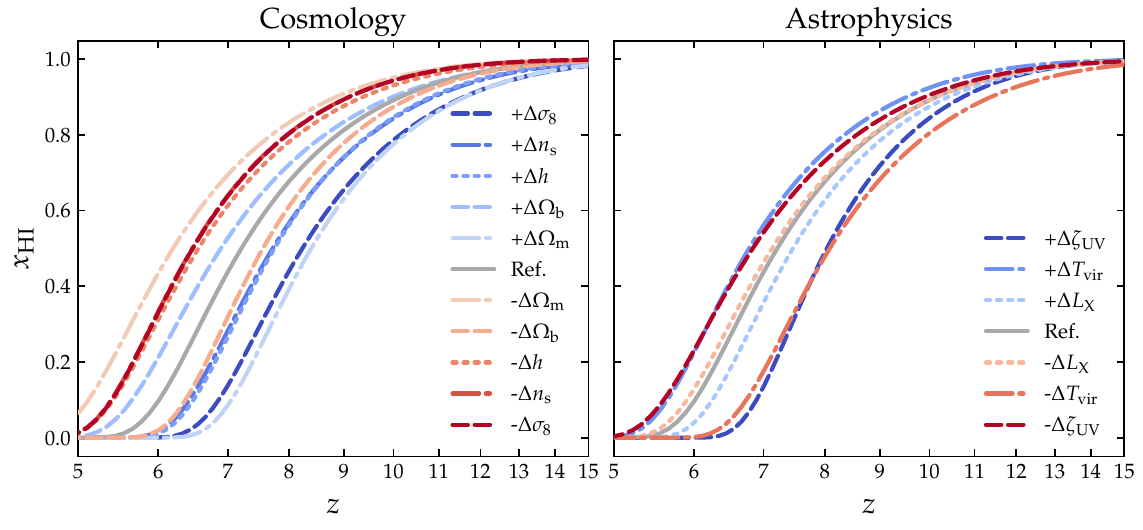}
    \caption{Neutral hydrogen history under variations of cosmological and astrophysical parameters. The reference model (``Ref.'') depicts the best-fit reionization history constrained by CMB + lensing  + DW. Using the reference model as a base, we then increase/decrease the value of a given parameter while leaving all others fixed. The increments and decrements are representative of the range sampled in our {\sc 21cmFAST} simulations. See Table \ref{tab:trends} for the specific parameter shifts and their corresponding impact on the midpoint of reionization.}
    \label{fig:trends}
\end{figure}

Eqs.~(\ref{eq:pivot}, \ref{eq:tilt}) establish a map between cosmological parameters and reionization history, a connection often overlooked in CMB analyses. Figure \ref{fig:trends} illustrates how cosmological and astrophysical parameters shape the evolution of our Robust Gompertzian reionization model.

\begin{table}
    \centering
    \caption{Parameters shifts used in Figure \ref{fig:trends}. The total shift is defined as $\Delta x \equiv x - x^{\rm Ref.}$, where the reference model corresponds to the best-fit from CMB + lensing + DW. The relative shift is given by $\Delta x / x^{\rm Ref.}$. We also tabulate the impact of the shifts on the midpoint of reionization, $z_\reio$, relative to the reference model. }
    \begin{tabular}{c|cc|cc}
    \toprule
    Parameter & Total shift & Relative shift & $z_\reio - z_\reio^{\rm Ref.}$ & $(z_\reio - z_\reio^{\rm Ref.})/z_\reio^{\rm Ref.}$  \\
    \hline 
    $+\Delta \sigma_8$ & 0.089 & 0.109 & 1.02 & 0.141 \\
    $-\Delta \sigma_8$ & $-0.071$ & $-0.088$ & $-0.73$ & $-0.102$ \\ 
    $+\Delta \ns$ & 0.030 & 0.031 & 0.51 & 0.071 \\
    $-\Delta\ns$ & $-0.050$ & $-0.051$ & $-0.74$ & $-0.102$ \\ 
    $+\Delta h$ & 0.054 & 0.081 & 0.54 & 0.075 \\ 
    $-\Delta h$ & $-0.065$ & $-0.097$ & $-0.65$ & $-0.090$ \\
    $+\Delta\Omegab$ & 0.011 & 0.217 & $-0.32$ & $-0.044$ \\
    $-\Delta\Omegab$ & $-0.009$ & $-0.188$ & 0.35 & 0.048 \\
    $+\Delta\Omegam$ & 0.087 & 0.279 & 1.18 & 0.163 \\ 
    $-\Delta\Omegam$ & $-0.073$ & $-0.232$ & $-1.01$ & $-0.140$ \\
    \hline
    \multicolumn{5}{c}{Astrophysics}\\
    \hline
    $+\Delta\zetaUV$ & 10.8 & 0.446 & 0.82 & 0.114 \\
    $-\Delta\zetaUV$ & $-4.20$ & $-0.173$ & $-0.38$ & $-0.052$ \\
    $+\Delta\Tvir$ & 0.14 & 0.031 & $-0.43$ & $-0.060$ \\
    $-\Delta\Tvir$ & $-0.026$ & $-0.057$ & 0.87 & 0.121 \\
    $+\Delta\LX$ & 0.80 & 0.020 & 0.25 & 0.035 \\
    $-\Delta\LX$ & $-1.20$ & $-0.030$ & $-0.082$ & $-0.011$\\
    \bottomrule
    \end{tabular}
    \label{tab:trends}
\end{table}

Figure \ref{fig:trends} -- and Table \ref{tab:trends} -- shows how parameters that enhance structure formation hasten reionization, as can be seen for higher values of $\sigma_8$ and $\Omegam$. Since our {\sc 21cmFAST} simulations assume that faint galaxies are the primary drivers of reionization, it is unsurprising that larger $\ns$, which boosts power on small scales, leads to a faster reionization process. Notably, higher $\Omegab$ delays reionization likely due to increased neutral hydrogen content in the IGM requiring ionization. The Hubble parameter $h$ exhibits competing effects: while higher values increase physical densities (potentially delaying reionization similar to $\Omegab$), they can also simultaneously enhance halo collapse fractions and hence source formation. The net acceleration observed at higher $h$ indicates that structure formation enhancement dominates over other effects.

The right panel in Figure \ref{fig:trends} indicates how astrophysical parameters govern reionization timing. For instance, a reduced ionizing photon budget significantly delays the reionization process, whether from suppressed escape fractions (dashed dark red line) or diminished UV source populations (dash-dotted blue curve). Furthermore, enhanced X-ray heating (dotted light blue) moderately accelerates reionization.

From Table \ref{tab:trends}, $\ns$ and $\Tvir$ are the most dominant parameters in the model since small percentage shifts translate to significant percentage shifts in the midpoint of reionization. In contrast, $\Omegab$ would be the less sensitive as large percentage shifts yield small percentage shifts in $z_\reio$. The differences in the level of sensitivity to reionization does not indicate that a parameter does not play an important role, e.g. there would not be any reionization without baryons, but that the more \emph{economical} way to change the timeline of reionization would be to shift the tilt of the primordial power spectrum instead of $\Omegab$. 

\section{Overfitting: alternative mapping for the Rgomp model}
\label{app:overfit}
In \S\ref{sec:gomp} we use our 1024 {\sc 21cmFAST} realizations to obtain a set of input pivots and tilts accompanied by the corresponding labels, i.e. the cosmological and astrophysical parameters used in the simulations. We use all of our $x_\HI$ samples to obtain Eqs.~(\ref{eq:pivot}, \ref{eq:tilt}). Here we instead use half of our exterior and inner sets to train alternative expressions for the mapping between cosmology and reionization. Our alternative expressions for the pivot and tilt are given by
\begin{align}
    &\ln \alpha(\boldsymbol{\theta}) = \left(\frac{\Omegab}{\sigma_8 \Omegam}\right)^{\sigma_8 h} - \left( \big( \sigma_8 (19.92 + \zetaUV + 0.08891^{(40.43 - \LX)}) \big)^{\ns/\Tvir} + \Omegam \right)\, , \label{eq:SRH_pivot} \\
    &\beta(\boldsymbol{\theta}) = \ns^{-1} \left( \frac{\zetaUV}{h} (0.1254 - \Omegab) - \frac{\sigma_8}{\Tvir \Omegam} \right) + e^{0.03961 \LX}\, . \label{eq:SRH_tilt}
\end{align}
{\sc PySR} assigns complexity, i.e. a numerical value, to operators and variables, essentially assigning some cost to penalize more complex functions. Expressions \ref{eq:SRH_pivot} and \ref{eq:SRH_tilt} correspond to complexity 27 and 20, respectively. Furthermore, their training losses are $5.50 \times 10^{-5}$ and $9.41 \times 10^{-3}$. In comparison, the main analysis pivot -- Eq.~(\ref{eq:pivot}) -- has complexity 34 and a loss of $2.46 \times 10^{-5}$ while the tilt has complexity 24 and loss $6.81 \times 10^{-3}$. Hence, Eq.~(\ref{eq:SRH_pivot}, \ref{eq:SRH_tilt}) are slightly less complex than the main analysis, with a slightly degraded performance training-loss wise.

Armed with Eq.~(\ref{eq:SRH_pivot}, \ref{eq:SRH_tilt}), we can compute the validation loss using the remaining half of the data. We obtain $6.13 \times 10^{-5}$ and $0.0109$ for the pivot and tilt, respectively. The relatively similar performance in the training and validation loss indicates that we are likely safe from overfitting.

\section{A tangent too far: Hyperbolic tangent reionization model and astrophysical data}
\label{app:tanh}
Some readers might perceive the results in \S\ref{sec:val} as favoring the Gompertz-like reionization model, since incorporating astrophysical data tightens constraints on cosmological parameters. We emphasize that this choice stems from the inability of the $\tanh$ model to adequately fit reionization data. To explicitly demonstrate this limitation, here we present the $\tanh$ model failure to jointly fit CMB [\S\ref{ssec:cmb}] + lensing [\S\ref{ssec:cmb}] + BAO [\S\ref{ssec:lowz}] + reionization constraints [\S\ref{ssec:astro}]. 

\begin{figure}
    \centering
    \includegraphics[width=1.1\linewidth]{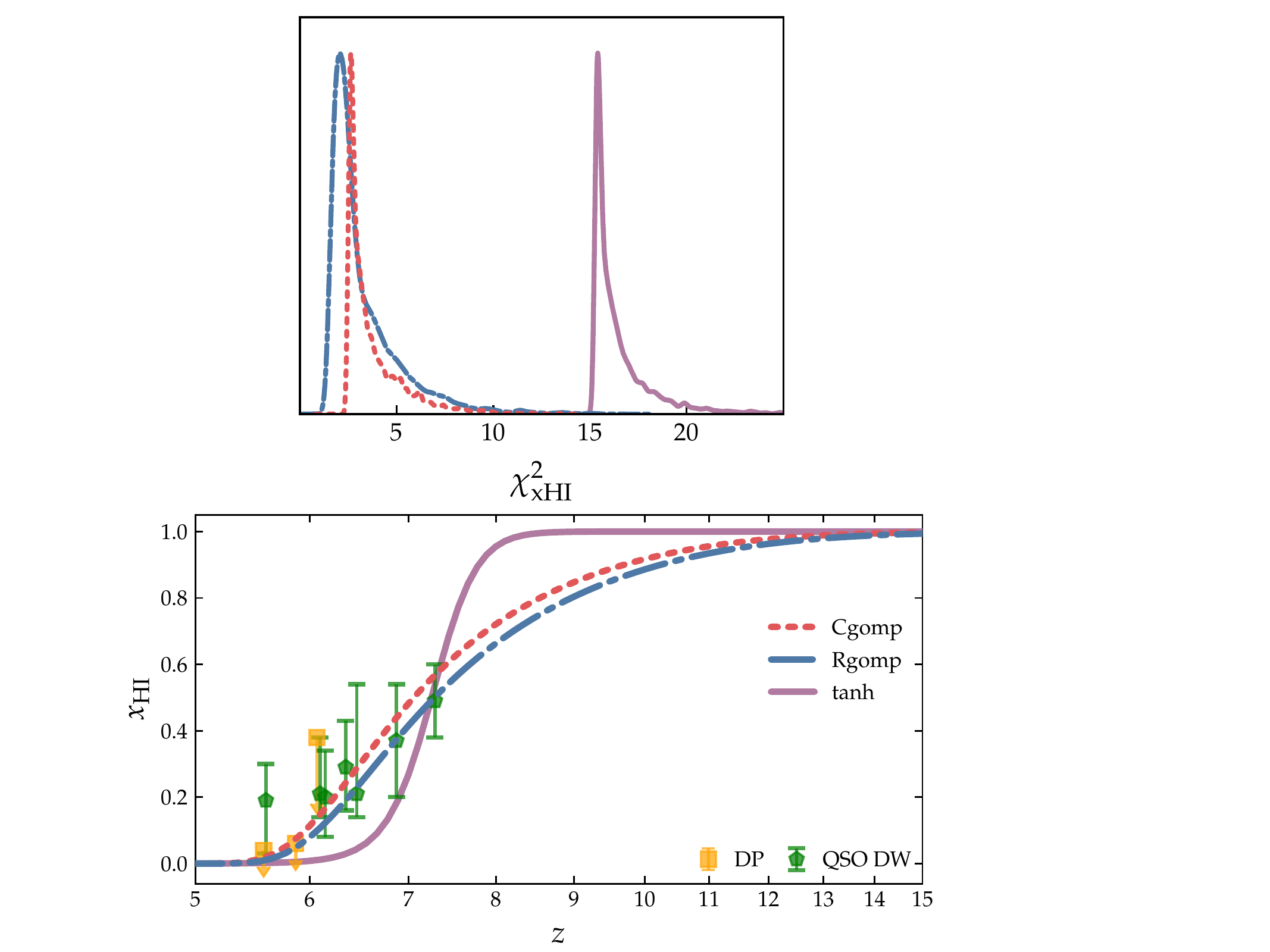}
    \caption{Inability of the $\tanh$ model to fit astrophysical constraints on the timeline of reionization. \textit{Top panel}: $\chi^2$ distributions from MCMC runs combining CMB + lensing + BAO + DW dataset, comparing three reionization models: Cgomp (dotted red line), Rgomp (dash-dotted blue), and $\tanh$ (solid purple). \textit{Bottom panel}: Reionization histories versus redshift for the same models, including quasar damping wing constraints (\S\ref{ssec:astro}). We also include the dark pixel (DP) constraints for reference. The $\tanh$ parametrization shows clear tension with astrophysical constraints while attempting to simultaneously fit CMB observations.}
    \label{fig:tanh_fail}
\end{figure}

Figure \ref{fig:tanh_fail} reveals the limitation of the $\tanh$ reionization model in simultaneously fitting CMB data and reionization observations. This failure originates from the symmetric nature of the hyperbolic tangent sigmoid model, which conflicts with the asymmetric physics that govern cosmic reionization. Cosmic reionization begins gradually at high redshifts due to sparse ionizing sources, then accelerates during intermediate stages as source formation increases and ionized bubbles coalesce. The symmetric $\tanh$ shape cannot capture this characteristic asymmetry, leading to biased constraints on the reionization timeline. 

We emphasize that even if one allows for the duration of reionization to be a free parameter, the $\tanh$ model still has a poor performance in the joint fit.

\section{Impact of neutrino masses in the reionization history}
\label{app:Mnu}
The {\sc 21cmFAST} simulation suite used in \S\ref{sec:gomp} to obtain our new reionization model does not span different values of $\sum m_\nu$. Instead, they assume $\sum m_\nu = 0$ eV, i.e. no contribution from massive neutrinos to the total matter density budget. Here we justify this modeling choice by assessing the deviations of the neutral hydrogen fraction for models with different $\sum m_\nu$.

\begin{figure}
    \centering
    \includegraphics[width=0.7\linewidth]{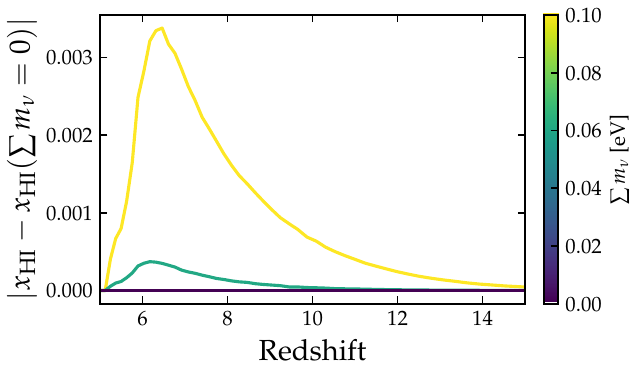}
    \caption{Impact of the massive neutrinos in the reionization history. The colored lines showcase the absolute difference between $x_\HI$ for the corresponding $\sum m_\nu$ shown in the color bar. }
    \label{fig:Mnu_xHI}
\end{figure}

Figure \ref{fig:Mnu_xHI} indicates that for a $\sum m_\nu \approx 0.10$ eV the maximum deviation with respect to a reionization history obtained from $\sum m_\nu = 0$ eV is roughly 0.3\%. Thus, we deemed this impact sufficiently small to safely use our Rgomp model for analyses with varying $\sum m_\nu$. 

\section{Transition between matter domination and evolving dark energy}
\label{app:w0wa}
An overly cautious read of the best fit for $w(z)$ in the Rgomp model may generate concerns about our usage of this reionization model to analyze $w_0w_a$CDM scenarios given that the {\sc 21cmFAST} simulations used to construct our model assume $\Lambda$CDM. This choice is justified since reionization ends well into the matter-domination era. Here we just briefly quantify when the transition occurs in $w_0w_a$CDM for a few representative cosmologies by looking at the epoch of equality between matter and dark energy. 

The evolving dark energy equation of state parameter for the Chevallier-Polarski-Linder parametrization is given by 
$w(a) = w_0 + w_a (1 - a)$ \cite{2001IJMPD..10..213C,2003PhRvL..90i1301L}. Plugging this into the continuity equation we obtain
\begin{equation}
    \label{eq:rhoDE}
    \rho_{\rm DE} (a) = \rho_{\mathrm{DE},0} \,a^{-3(1 + w_0 + w_a)} \, e^{3 w_a (a - 1)} \, ,
\end{equation}
where $\rho_{\mathrm{DE},0} = \Omega_{\rm DE} \rho_{\rm crit}$ with $\rho_{\rm crit}$ the usual critical density. 

Equating Eq.~(\ref{eq:rhoDE}) with the analog matter contribution we can obtain the scale factor of equality ($a^{\rm m, DE}_{\rm eq})$ between the two components, i.e.
\begin{equation}
    \label{eq:equality}
    \left(a^{\rm m, DE}_{\rm eq}\right)^{3 (w_0 + w_a)} e^{-3 w_a (a^{\rm m, DE}_{\rm eq} - 1)} = \frac{\Omega_{\rm DE}}{\Omegam} \, .
\end{equation}
Given $w_0$, $w_a$, and $\Omegam$ we can solve Eq.~(\ref{eq:equality}) using bisection and obtain the corresponding redshift of equality. We tabulate our findings for a few parameter combinations in Table \ref{tab:equality}.

\begin{table}
    \centering
    \begin{tabular}{l|l|c}
    \toprule
     Source & Data & $z^{\rm m, DE}_{\rm eq}$ \\
     \hline
     DESI DR2 & DESI + CMB & 0.472 \\
     DESI DR2 & DESI + CMB + DESY5 &  0.374 \\
     Rgomp & CMB + BAO + lensing + DW & 0.359\\
     \bottomrule
    \end{tabular}
    \caption{Redshift of equality between matter and evolving dark energy. For values corresponding to DESI DR2 results, we have opted to keep the nomenclature used in \cite{2025arXiv250314738D}, i.e. DESI + CMB does include CMB lensing. Likewise, we have kept the nomenclature used in this work for the Rgomp entry. }
    \label{tab:equality}
\end{table}

\acknowledgments
This project is supported by the Basic and Frontier Research Project of PCL (grant No.~2025QYB012) and the Major Key project of Peng Cheng Laboratory. We acknowledge the Tsinghua Astrophysics High-Performance Computing platform at Tsinghua University, PCL's Cloud Brain, and the Materials Science and Engineering Research Center High-Performance Computing platform, CICIMA HPC, at University of Costa Rica for providing computational and data storage resources that have contributed to the research results reported within this paper.

PR acknowledges support from the Spanish Ministerio de Ciencia, Innovación y Universidades, through projects PID2022-138896NB; and the programme Unidad de Excelencia María de Maeztu, project CEX2020-001058-M

% Bibliography

%% [A] Recommended: using JHEP.bst file
\bibliographystyle{JHEP}
\bibliography{gomp.bib}
\end{document}